\documentclass[11pt,a4paper]{article}
\pdfoutput=1
\usepackage{jheppub}

\usepackage{multirow, graphicx,amssymb,url,mathrsfs,amsmath}
\usepackage{wrapfig,boxedminipage,setspace,subfigure,epsfig}
\usepackage{amsxtra,amstext,latexsym,dsfont,amsfonts}
\usepackage{color}
\usepackage[dvipsnames]{xcolor}
\usepackage{float}
\usepackage{slashed,comment}
\usepackage{contour,amsthm}

\usepackage{tikz}
\usetikzlibrary{calc,patterns,angles,quotes}
\usetikzlibrary{decorations.pathreplacing,decorations.markings,snakes}

\usepackage{tabularx, array}
\newcolumntype{L}[1]{>{\raggedright\arraybackslash}p{#1}}
\newcolumntype{C}[1]{>{\centering\arraybackslash}p{#1}}
\newcolumntype{R}[1]{>{\raggedleft\arraybackslash}p{#1}}

\newcommand{\pd}{\partial}
\newcommand{\Lk}{\mathbb{L}}
\newcommand{\cD}{\mathcal{D}}
\newcommand{\cB}{\mathcal{B}}
\newcommand{\Dser}[1]{\mathrm{D}^{+}_{#1}}
\newcommand{\ket}[1]{|#1)}
\newcommand{\bra}[1]{(#1|}
\newcommand{\SL}{\mathrm{SL}(2,\mathbb{R})}
\newtheorem{proposition}{Proposition}
\newtheorem{corollary}{Corollary}

\title{On the near-horizon interpretation of Krylov dynamics}

\author[a,b]{Hyun-Sik Jeong\,}

\emailAdd{hyunsik.jeong@apctp.org}

\preprint{\texttt{APCTP Pre2026 - 003}}

\affiliation[a]{Asia Pacific Center for Theoretical Physics, Pohang 37673, Korea}
\affiliation[b]{Department of Physics, Pohang University of Science and Technology, Pohang 37673, Korea}

\abstract{
The Lanczos recursion maps operator growth onto the motion of a wave function on a semi-infinite chain, and for linearly growing Lanczos coefficients the deep region of the chain behaves like the near-horizon region of a black hole. We develop this correspondence in the time domain and determine how far it extends. Writing the Krylov equation as an exactly skew-adjoint difference operator, we derive a unitary continuum limit whose corrections are suppressed by $n^{-2}$ once the origin of the chain is fixed by the offset of the Lanczos coefficients. For $b_n\simeq\alpha n$ the Krylov evolution is chiral transport in $\ln n$, which coincides with infalling waves near the horizon of an AdS$_2$ black hole when the Krylov depth is identified with the blueshift of an infalling particle; this fixes $\alpha=\pi T$, while staggered configurations of the chain carry the outgoing chirality. For $b_n=\alpha\sqrt{n(n+2h-1)}$ the correspondence is exact: the Krylov basis is the discrete series of $\mathrm{SL}(2,\mathbb{R})$ carried by a bulk scalar of mass $m^2L^2=h(h-1)$, the Liouvillian is the boost generator of the black hole at temperature $T=\alpha/\pi$, and positivity of $b_n^2$ reproduces the Breitenlohner--Freedman bound, saturated by strictly linear coefficients. Beyond leading order we prove that, for every Lanczos sequence with a smooth large-$n$ expansion, the ingoing solutions of the Krylov equation have coefficients polynomial in frequency to all orders in $1/n$, whereas a massive bulk field has a pole at the first special point $\omega=-2\pi iT$ with residue proportional to $m^2$. Smooth subleading Lanczos data therefore cannot encode the bulk mass locally. The obstruction persists for the BTZ black hole, for which we obtain the Lanczos coefficients in closed form.
}

\begin{document}
\maketitle

%
\section{Introduction}\label{sec:intro}
Krylov subspace methods have long been recognized as powerful tools for probing the dynamics of many-body systems~\cite{liesen2012krylov,Viswanath_1994}. In quantum physics, the Lanczos algorithm~\cite{krylov1931,lanczos1950} constructs an orthonormal basis of the operator subspace explored by Heisenberg evolution, and maps the growth of an operator onto the motion of a wave function on a semi-infinite one-dimensional chain~\cite{Mattis1981,Viswanath_1994}. The framework has recently experienced a resurgence of interest, driven by its use in characterizing operator growth, quantum complexity and chaos~\cite{Parker:2018yvk,Dymarsky:2019elm,Balasubramanian:2022tpr,Caputa:2024vrn,Nandy:2024evd,Rabinovici:2025otw,Jeong:2026gdc}, in quantum algorithms~\cite{Bharti:2021aa,Cortes:2022aa,Kirby2023exactefficient}, and in both unitary and open quantum dynamics~\cite{Bhattacharya2022,Liu:2022god,Bhattacharya2023,Baggioli:2025knt}.

In this language the dynamics is encoded in the Lanczos coefficients $b_n$, the hopping amplitudes of the chain. The universal operator growth hypothesis~\cite{Parker:2018yvk} states that $b_n$ grows asymptotically linearly, $b_n\simeq\alpha n$, in chaotic systems. Krylov complexity, the mean position on the chain, then grows as $e^{2\alpha t}$ and bounds the Lyapunov exponent of out-of-time-ordered correlators, $\lambda_L\le2\alpha$~\cite{Parker:2018yvk}. With the Wightman inner product at temperature $T$ one has, moreover, $\alpha\le\pi T$~\cite{Parker:2018yvk,Avdoshkin:2019trj}, so that this bound is at least as strong as the bound on chaos, $\lambda_L\le2\pi T$~\cite{Maldacena:2015waa}. The saturation $\alpha=\pi T$ is, however, not a diagnostic of maximal chaos: it holds kinematically for thermal correlators of conformal field theories (CFTs)~\cite{Dymarsky:2021bjq}. Within holography (AdS/CFT)~\cite{Maldacena:1997re,Gubser:1998bc,Witten:1998qj}, Krylov complexity has been related to the length of the Einstein-Rosen bridge in Jackiw-Teitelboim (JT) gravity and in the double-scaled SYK model~\cite{Jian:2020qpp,Lin:2022rbf,Rabinovici:2023yex,Heller:2024ldz,Ambrosini:2024sre,Fu:2025kkh}, in line with the broader program that relates quantum complexity to black-hole interiors~\cite{Susskind:2014rva,Stanford:2014jda,Chapman:2021jbh,Chen:2021lnq} and with the view of black holes as fast scramblers~\cite{Sekino:2008he,Maldacena:2015waa}.

A more refined question is whether the Krylov dynamical equation itself, and not only integrated quantities such as Krylov complexity, has a gravitational counterpart. Microscopic measures of operator growth are expected to encode near-horizon physics through the relation between operator size and the momentum of an infalling particle~\cite{Susskind:2018tei,Brown:2018kvn,Lin:2019qwu,Magan:2018nmu,Kar:2021nbm,Iliesiu:2021ari}. Recently, Dodelson~\cite{Dodelson:2025rng} showed that the Lanczos recursion can be recast as a discrete scattering problem with the structure of wave propagation on a black-hole background. In the continuum limit, with $dz=dn/2b_n$ and $\psi=\sqrt{b_n}\,\phi$, linear growth $b_n\simeq(\pi/\beta_0)n$ yields a free wave equation in a tortoise-like coordinate $z\simeq(\beta_0/2\pi)\ln n$; matching to the tortoise coordinate of a black hole of inverse temperature $\beta_0$ identifies large $n$ with the near-horizon region, $n\sim(1-r_h/r)^{-1/2}$; and subleading terms in $b_n$ define a scattering potential, whose smooth and staggered parts have different analytic consequences for the Green's function~\cite{Dodelson:2025rng}.

\begin{figure}[t]
\centering
\includegraphics[width=0.72\textwidth]{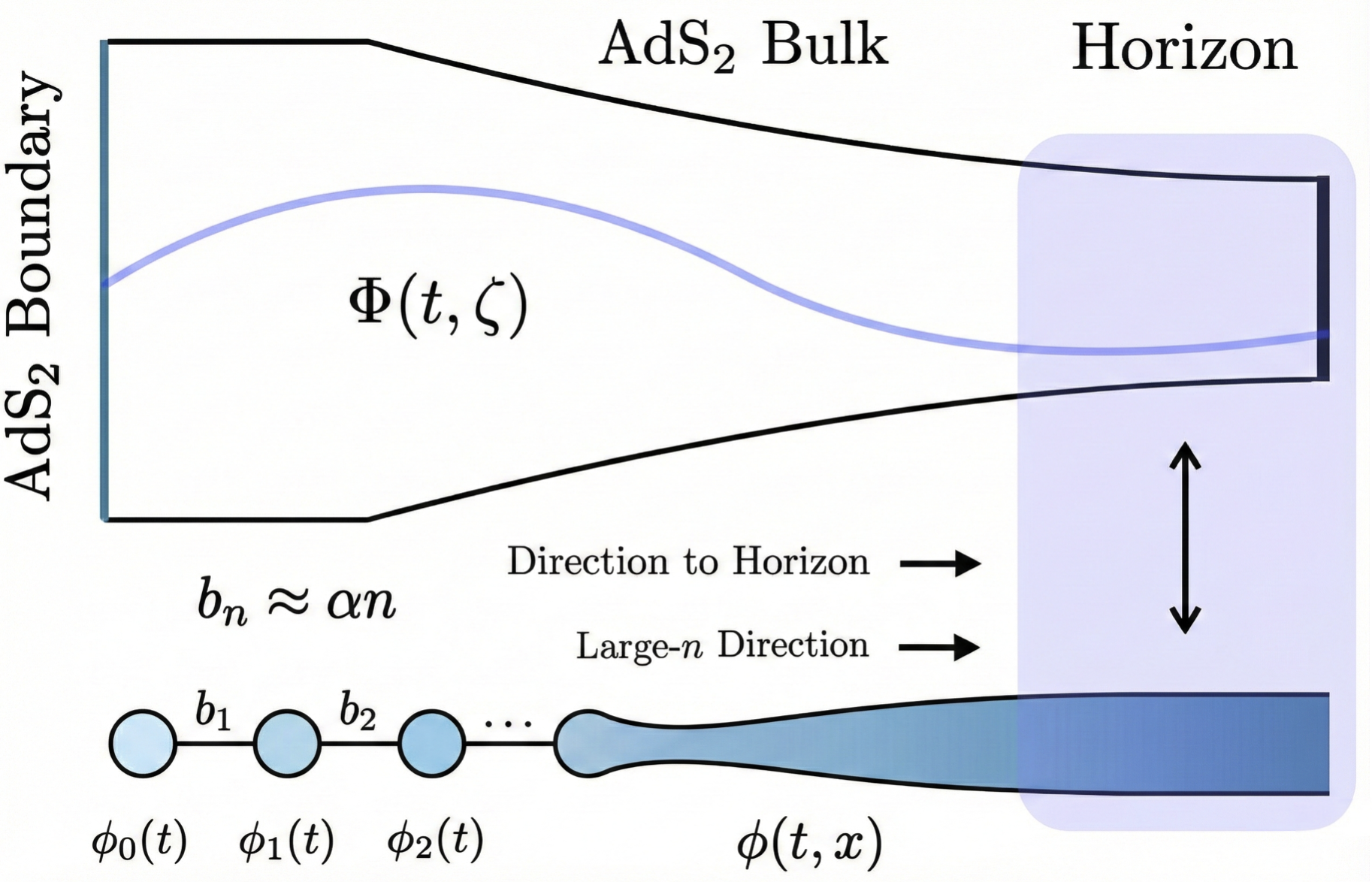}
\caption{Schematic of the correspondence. The deep Krylov region ($n\gg1$), where the continuum field $\phi(t,x)$ is valid, maps onto the near-horizon AdS$_2$ throat, where a field $\Phi(t,\zeta)$ falls toward the horizon.}
\label{fig:schematic}
\end{figure}

The purpose of this paper is to develop this correspondence in the time domain, to make it exact when both sides carry the same symmetry, and to determine how far a local identification of the deep Krylov chain with the near-horizon region can be pushed (figure~\ref{fig:schematic}). Our main results are as follows.
\begin{enumerate}
\item \textit{Unitary continuum limit} (section~\ref{sec:continuum}). Attaching the Lanczos coefficients to the links of the chain, we write the Krylov equation exactly as a skew-adjoint difference operator whose derivative expansion is skew-adjoint order by order. The leading term is chiral transport, the time-domain counterpart of the free wave equation of ref.~\cite{Dodelson:2025rng}. The corrections are of relative order $n^{-2}$ when the origin of the continuum coordinate is fixed by the constant offset of $b_n$, and of order $n^{-1}$ otherwise; a naive truncation of the derivative expansion violates unitarity.
\item \textit{Leading-order dictionary} (section~\ref{sec:leading}). For $b_n\simeq\alpha n$ the Krylov evolution is rigid chiral transport in $\ln n$, which coincides with infalling waves near the horizon of an AdS$_2$ black hole; the staggered sector of the chain carries the outgoing chirality. The leading-order matching by itself fixes only the relative normalization of two radial coordinates. The value $\alpha=\pi T$ follows once the Krylov depth is identified with the blueshift of an infalling particle, an identification equivalent to that of ref.~\cite{Dodelson:2025rng}, which for a general static metric reads $b(n)=\frac14 n f'(r)$, where $f(r)$ is the blackening factor.
\item \textit{Exact $\SL$ dictionary} (section~\ref{sec:sl2}). For $b_n=\alpha\sqrt{n(n+2h-1)}$ the correspondence holds without any continuum limit. The Krylov basis is the discrete series $\Dser{h}$ carried by a bulk scalar of mass $m^2L^2=h(h-1)$, the Liouvillian is $2\pi T$ times the boost Killing vector of the AdS$_2$ black hole, and $\alpha=\pi T$ follows from the normalization of non-compact generators, consistently with the Kubo-Martin-Schwinger (KMS) strip~\cite{Kubo_1957,Martin:1959aa}. Krylov complexity is the excess global energy of the boosted bulk particle, the late-time decay at the edge of the chain is set by the lowest quasinormal mode, and positivity of $b_n^2$ reproduces the Breitenlohner--Freedman (BF) bound~\cite{Breitenlohner:1982bm,Breitenlohner:1982jf}, saturated by strictly linear coefficients.
\item \textit{No local matching of the mass} (section~\ref{sec:nogo}). Subleading Lanczos data enter the deep-Krylov equation at the same order as the leading near-horizon corrections of the Klein--Gordon equation, and must be kept. We identify the invariant content of these corrections under local field redefinitions and coordinate changes, and prove that for every Lanczos sequence with a smooth large-$n$ expansion the ingoing solutions of the Krylov equation have coefficients polynomial in frequency to all orders in $1/n$. A massive bulk field instead has a pole at the first special point $\omega=-2\pi iT$~\cite{Blake:2019otz}, with residue proportional to $m^2$. Smooth subleading Lanczos data therefore cannot encode the bulk mass locally, at any order. This extends to all orders the observation of ref.~\cite{Dodelson:2025rng} that smooth corrections produce entire functions of the frequency, and shows that the mass is encoded at the edge of the chain.
\item \textit{Beyond AdS$_2$} (section~\ref{sec:beyond}). For the BTZ black hole at momentum $q$ the Lanczos coefficients follow in closed form from the continuous Hahn polynomials; they are smooth, and the bulk residue at the first special point, $\propto q^2/(2\pi T)^2+(\Delta-1)^2$, vanishes only at the pole-skipping momenta~\cite{Blake:2019otz,Blake:2017ris,Grozdanov:2017ajz,Blake:2018leo,Ahn:2025qat}. The no-go therefore extends to BTZ at every real momentum (except $q=0$ for $\Delta=1$), including massless fields. Holographic Wightman densities have no zeros, because the zeros of the thermal factor cancel the special-point poles of the horizon-normalized bulk solution; the near-horizon residues are encoded in the spectral factor of the density, a non-local functional of the chain. 
\item \textit{Power-law growth} (section~\ref{sec:power}). For $b_n\propto n^p$ with $p\ge0$ the same dictionary yields a regular horizon only for $p=1$ (non-extremal) and $p=0$ (extremal). For $p>1$ the horizon lies at finite tortoise distance.
\end{enumerate}
Section~\ref{sec:discussion} summarizes the dictionary, its scope and its testable predictions. Technical derivations are collected in appendices~\ref{app:expansion}--\ref{app:btz}.

%
\section{Krylov dynamics}\label{sec:krylov}

\subsection{Lanczos algorithm and the Krylov chain}
The Lanczos approach~\cite{liesen2012krylov,Viswanath_1994,krylov1931,lanczos1950} to operator dynamics starts from the Heisenberg evolution $\pd_tO(t)=i\mathcal{L}O(t)$, where $\mathcal{L}=[H,\cdot\,]$ is the Liouvillian. The evolution $O(t)=e^{i\mathcal{L}t}O(0)$ is confined to the Krylov subspace ${\rm span}\{\mathcal{L}^nO(0)\}_{n\ge0}$. Representing $O$ as a state $\ket{O}$ in operator space, with an inner product $\bra{A}B)$ for which $\mathcal{L}$ is Hermitian, an orthonormal Krylov basis $\{\ket{O_n}\}$ is generated by
\begin{equation}
\ket{A_n}=\mathcal{L}\ket{O_{n-1}}-b_{n-1}\ket{O_{n-2}},\qquad \ket{O_n}=b_n^{-1}\ket{A_n},\qquad b_n=\bra{A_n}A_n)^{1/2},
\label{eq:lanczos}
\end{equation}
with $\ket{O_0}=\ket{O(0)}$ and $b_0=0$. At infinite temperature one uses $\bra{A}B)={\rm tr}[A^\dagger B]/{\rm tr}\,I$. At temperature $T=1/\beta$ we use the Wightman inner product $\bra{A}B)_\beta={\rm tr}[\rho^{1/2}A^\dagger\rho^{1/2}B]$, $\rho=e^{-\beta H}/Z$, which is the one relevant for the bound on chaos~\cite{Parker:2018yvk}. Expanding $\ket{O(t)}=\sum_n i^n\phi_n(t)\ket{O_n}$ gives the Krylov equation on a semi-infinite chain,
\begin{equation}
\pd_t\phi_n(t)=b_n\,\phi_{n-1}(t)-b_{n+1}\,\phi_{n+1}(t),\qquad n\ge0,
\label{eq:chain}
\end{equation}
with $\phi_n(0)=\delta_{n0}$, real $\phi_n(t)$, and $\sum_n\phi_n^2=1$ conserved. The Lanczos coefficients are determined by the autocorrelation function $\phi_0(t)=\bra{O_0}O(t))$, equivalently by its spectral density, and their asymptotics reflect the analytic structure of $\phi_0(t)$: linear growth with rate $\alpha$ corresponds to a singularity of $\phi_0(t)$ at $|{\rm Im}\,t|=\pi/2\alpha$, or to a spectral density decaying as $e^{-\pi|\omega|/2\alpha}$~\cite{Parker:2018yvk,Avdoshkin:2019trj,Dodelson:2025rng}.

The operator growth hypothesis~\cite{Parker:2018yvk} states that $b_n\simeq\alpha n$ for chaotic lattice systems in more than one spatial dimension (in one dimension $b_n\sim n/\ln n$) and for non-local models such as SYK. Then Krylov complexity $K(t)=\sum_n n\,\phi_n^2$ grows as $e^{2\alpha t}$, and $\lambda_L\le2\alpha$~\cite{Parker:2018yvk}. Since the Wightman autocorrelation is analytic in the strip $|{\rm Im}\,t|<\beta/2$, one has $\alpha\le\pi T$~\cite{Parker:2018yvk,Avdoshkin:2019trj}. Saturation $\alpha=\pi T$ occurs, for instance, in the low-temperature limit of large-$q$ SYK~\cite{Parker:2018yvk} and for conformal two-point functions~\cite{Dymarsky:2021bjq,Parker:2018yvk}; it is a statement about Krylov growth and does not by itself imply $\lambda_L=2\pi T$~\cite{Dymarsky:2021bjq}.

\subsection{The \texorpdfstring{$\SL$}{SL(2,R)} chain}\label{sec:sl2chain}
A central example is
\begin{equation}
b_n=\alpha\sqrt{n(n+2h-1)},\qquad h>0,
\label{eq:bnsl2}
\end{equation}
for which the Krylov equation is solved exactly by~\cite{Parker:2018yvk,Caputa:2021sib}
\begin{equation}
\phi_n(t)=\sqrt{\frac{(2h)_n}{n!}}\;\frac{\tanh^n(\alpha t)}{\cosh^{2h}(\alpha t)},\qquad K(t)=2h\sinh^2(\alpha t),
\label{eq:sl2sol}
\end{equation}
where $(2h)_n=\Gamma(2h+n)/\Gamma(2h)$. The autocorrelation is $\phi_0(t)=\cosh^{-2h}(\alpha t)$. Conversely, the Wightman-ordered thermal two-point function of an operator of dimension $\Delta$ in a conformal quantum mechanics at temperature $T$, for instance the boundary dual of a probe scalar in an AdS$_2$ black hole~\cite{Maldacena:2016hyu,Spradlin:1999bn}, is proportional to $[\pi T/\cosh(\pi Tt)]^{2\Delta}$; it generates the coefficients~\eqref{eq:bnsl2} with $\alpha=\pi T$ and $h=\Delta$~\cite{Parker:2018yvk,Dymarsky:2021bjq} (appendix~\ref{app:sl2}). For $h=\frac12$ the coefficients are exactly linear, $b_n=\alpha n$.

%
\section{Unitary continuum limit}\label{sec:continuum}

\begin{figure}[t]
\centering
\includegraphics[width=0.72\textwidth]{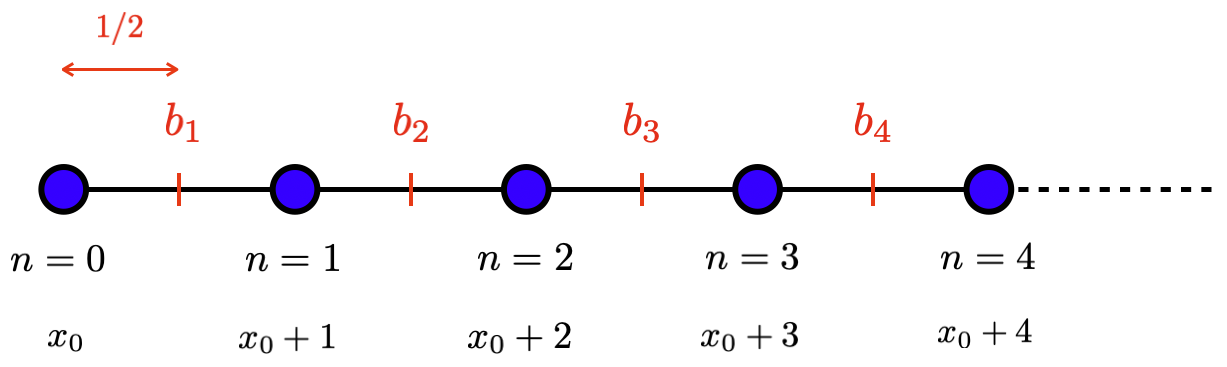}
\caption{Link-centered discretization of the Krylov chain. Site $n$ (blue dots) is placed at $x_n = n + x_0$. The Lanczos coefficient $b_n$, which couples sites $n-1$ and $n$, is attached to the midpoint of that link (red ticks), half a lattice spacing from each neighboring site, so that $b_n = B(x_n - \tfrac12)$ as in eq.~\eqref{eq:link}. The dashed line indicates that the chain is semi-infinite.}
\label{fig:schematic2}
\end{figure}

\subsection{Link-centered interpolation}
To describe the deep Krylov region we attach the chain to a continuous coordinate, with site $n$ located at $x_n=n+x_0$, and we attach each Lanczos coefficient to the midpoint of the link it multiplies (figure~\ref{fig:schematic2}),
\begin{equation}
b_n=B\big(x_n-\tfrac12\big),
\label{eq:link}
\end{equation}
with $B(x)$ smooth; since $b_n$ is defined only at integer $n$, a derivative expansion requires an interpolation $B(x)$. Writing $\phi_n(t)=\phi(t,x_n)$ for a smooth interpolating field, eq.~\eqref{eq:chain} is the restriction to $x=x_n$ of
\begin{equation}
\pd_t\phi=B\big(x-\tfrac12\big)\,\phi(x-1)-B\big(x+\tfrac12\big)\,\phi(x+1)\equiv\Lk\,\phi,\qquad
\Lk=e^{-\frac{1}{2}\pd_x}\,B\,e^{-\frac{1}{2}\pd_x}-e^{\frac{1}{2}\pd_x}\,B\,e^{\frac{1}{2}\pd_x}.
\label{eq:Lexact}
\end{equation}
Since $(e^{\pm\frac12\pd_x})^\dagger=e^{\mp\frac12\pd_x}$ on $L^2(\mathbb{R},dx)$, the operator $\Lk$ is manifestly skew-adjoint, $\Lk^{\dagger} = -\Lk$; $\int\phi^2dx$ is the continuum counterpart of $\sum_n\phi_n^2$. The link-centered (Weyl-ordered) form~\eqref{eq:Lexact} is equivalent to the site-centered form $\tilde Be^{-\pd_x}-e^{\pd_x}\tilde B$ with $\tilde B(x)=B(x-\frac12)$, but it is the natural one for the derivative expansion: each term of fixed total order is homogeneous in $x$ (for $B\propto x$, the terms with $2m+1$ derivatives are suppressed by $x^{-2m}$), whereas a site-centered expansion contains terms of relative order $x^{-1}$ that cancel only between different orders.

\subsection{Derivative expansion and unitarity}
Expanding eq.~\eqref{eq:Lexact} in derivatives (appendix~\ref{app:expansion}), every term of fixed total order in derivatives is separately skew-adjoint:
\begin{equation}
\Lk=-\big(2B\pd_x+B'\big)-\Big(\tfrac{1}{3}B\pd_x^3+\tfrac12B'\pd_x^2+\tfrac14B''\pd_x+\tfrac1{24}B'''\Big)+\dots
\label{eq:Lexp}
\end{equation}
For $b_n=\alpha n$ one has $x_0=\frac12$, $B=\alpha x$ and, in closed form, $\Lk=-\alpha\{x,\sinh\pd_x\}$. The second bracket in eq.~\eqref{eq:Lexp} must be kept or dropped as a whole. Keeping only its term $-\frac{\alpha}{2}\pd_x^2$, as a truncation of the Taylor expansion of eq.~\eqref{eq:chain} at second order would do, produces a backward-diffusion term and violates unitarity, $\frac{d}{dt}\int\phi^2dx=\alpha\int(\pd_x\phi)^2dx>0$; the resulting initial-value problem is ill-posed (appendix~\ref{app:truncation}). Since $B^{(j)}=O(x^{1-j})$, for a wave packet of width $\Delta x\sim x$ the second bracket is suppressed by $x^{-2}\sim n^{-2}$ relative to the first. To leading order,
\begin{equation}
\pd_t\phi=-\big(2B\pd_x+B'\big)\phi,
\label{eq:transport}
\end{equation}
which conserves $\int\phi^2dx$ exactly. With the Krylov tortoise coordinate and rescaled field
\begin{equation}
s(x)=\int^x\frac{dx'}{2B(x')},\qquad \chi=\sqrt{2B}\,\phi,
\label{eq:tortoiseK}
\end{equation}
eq.~\eqref{eq:transport} becomes $(\pd_t+\pd_s)\chi=0$: every profile is rigidly translated, $\chi=F(s-t)$, with conserved $\int\chi^2ds$. The coordinate $s$ and the field $\chi$ coincide, up to normalization, with the variables $z$ and $\psi$ of ref.~\cite{Dodelson:2025rng}, and eq.~\eqref{eq:transport} is the first-order, time-domain counterpart of the free wave equation obtained there in the frequency domain. Leading-order continuum approximations were also used to analyze Krylov complexity in refs.~\cite{Parker:2018yvk,Barbon:2019wsy,Rabinovici:2023yex}.

\subsection{Origin rule and logarithmic depth}\label{sec:origin}
For asymptotically linear coefficients we write
\begin{equation}
b_n=\alpha n+a_0+\frac{a_1}{n}+O(n^{-2}).
\label{eq:bnexp}
\end{equation}
Choosing the origin of the continuum coordinate as
\begin{equation}
x_0=\tfrac12+\frac{a_0}{\alpha},
\label{eq:originrule}
\end{equation}
the interpolating function~\eqref{eq:link} takes the form
\begin{equation}
B(x)=\alpha x+\frac{\beta_1}{x}+O(x^{-2}),\qquad \beta_1=a_1,
\label{eq:Bexp}
\end{equation}
without a constant term. With this choice, the corrections to eq.~\eqref{eq:transport} are suppressed by $x^{-2}$; with any other choice of origin a constant term in $B$ survives and the corrections are of relative order $x^{-1}$.

We introduce the logarithmic depth and the rescaled field
\begin{equation}
U=\ln x,\qquad \Phi(t,U)=x^{1/2}\phi(t,x),\qquad \int|\Phi|^2dU=\int\phi^2dx,
\label{eq:logdepth}
\end{equation}
and write $\cD\equiv\pd_U=x\pd_x$ acting on $\Phi$. In these variables the Krylov equation reads (appendix~\ref{app:logdepth})
\begin{equation}
\pd_t\Phi=-\alpha\Big[2\cD+e^{-2U}R(\cD)\Big]\Phi+O(e^{-3U}),\qquad
R(\cD)=(\cD-1)\Big[\tfrac13\big(\cD^2-2\cD+\tfrac34\big)+\frac{2\beta_1}{\alpha}\Big],
\label{eq:subleading}
\end{equation}
and, to leading order,
\begin{equation}
\pd_t\Phi=-2\alpha\,\cD\Phi,\qquad \pd_t^2\Phi=4\alpha^2\cD^2\Phi .
\label{eq:chiral}
\end{equation}
The correction $R(\cD)$ collects the two sources of $O(x^{-2})$ effects: the third-order derivative terms of eq.~\eqref{eq:Lexp}, and the subleading Lanczos coefficient $\beta_1$. For the $\SL$ coefficients~\eqref{eq:bnsl2}, $a_0=\alpha(h-\frac12)$, so that $x_0=h$, and $B=\alpha\sqrt{x^2-(h-\frac12)^2}$, so that $\beta_1=-\frac\alpha2(h-\frac12)^2$ and
\begin{equation}
R(\cD)=\tfrac13(\cD-1)\big(\cD(\cD-2)-3h(h-1)\big),
\label{eq:Rsl2}
\end{equation}
with the next corrections at $O(e^{-4U})$. The representation label enters only through the Casimir $h(h-1)$, at relative order $n^{-2}$, accompanied by $h$-independent terms of third order in $\cD$. More generally, at this order the deep-Krylov equation depends on the Lanczos data only through $\beta_1/\alpha$: every smooth chain is locally indistinguishable at $O(n^{-2})$ from an $\SL$ chain with effective Casimir $h_{\rm eff}(h_{\rm eff}-1)=-2\beta_1/\alpha-\frac14$. We return to the invariant content of $R(\cD)$ in section~\ref{sec:nogo}.

\subsection{Three consequences}\label{sec:consequences}

\paragraph{Rigid translation and Krylov complexity.} Deep in the chain the wave packet moves rigidly in $U=\ln x$ at speed $2\alpha$. Within eq.~\eqref{eq:chiral}, $\langle x\rangle(t)=e^{2\alpha t}\langle x\rangle(0)$ for \emph{any} profile, so $K(t)\propto e^{2\alpha t}$, the $O(n^{-2})$ corrections affecting only the prefactor.

\begin{figure}[t]
\centering
\includegraphics[width=6.2cm]{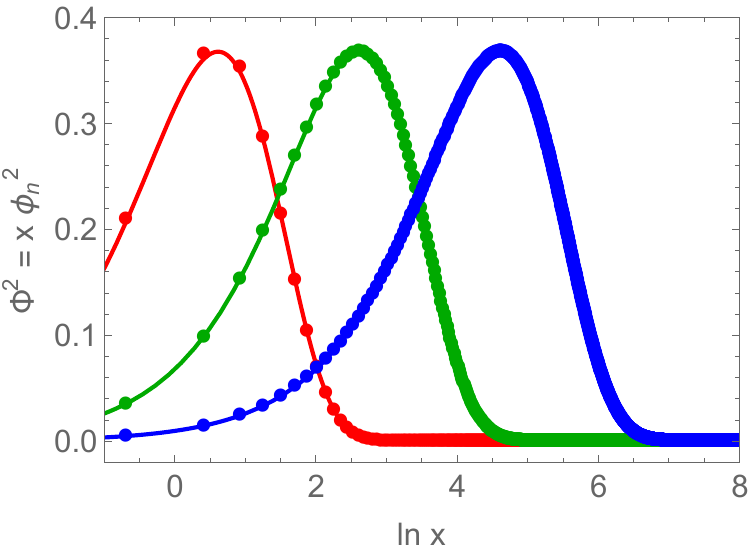}
\qquad
\includegraphics[width=6.4cm]{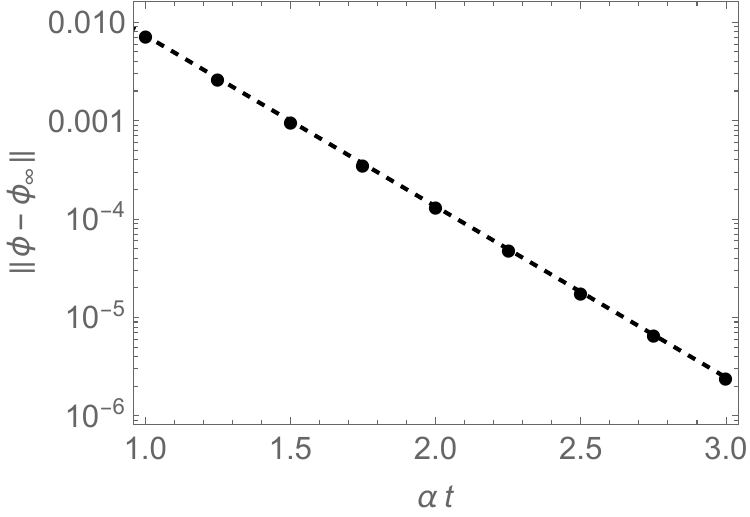}
\caption{Numerical test of the continuum limit for the $\SL$ chain with $h=\frac12$ ($b_n=\alpha n$), with the origin $x_n=n+\frac12$ of eq.~\eqref{eq:originrule}. \textit{Left:} Exact lattice solution~\eqref{eq:sl2sol}, $x_n\phi_n^2$ (dots), at $\alpha t=1,2,3$ (red, green, blue), against the rigid profile $\Phi_\infty^2(\ln x_n-2\alpha t)$ of eq.~\eqref{eq:profile} (lines). \textit{Right:} The $\ell^2$ deviation from the rigid profile, $\|\phi-\phi_\infty\|=\big[\sum_n\big(\phi_n(t)-\phi_{\infty,n}(t)\big)^2\big]^{1/2}$ with $\phi_{\infty,n}(t)=x_n^{-1/2}\,\Phi_\infty(\ln x_n-2\alpha t)$ (dots). The dashed line is $\propto e^{-4\alpha t}$, i.e.\ $\propto n^{-2}$ at the packet.}
\label{fig:numerics}
\end{figure}

\paragraph{Rigid profile.} For the $\SL$ chain the exact solution~\eqref{eq:sl2sol} approaches a rigidly translated profile,\footnote{For $b_n=\alpha\sqrt{n(n+2h-1)}$, $\phi_n=[\Gamma(n+2h)/(\Gamma(2h)n!)]^{1/2}\tanh^n(\alpha t)\cosh^{-2h}(\alpha t)$. With $x=n+h$, $\Gamma(n+2h)/\Gamma(n+1)=x^{2h-1}[1+O(x^{-2})]$, and $\ln\tanh\alpha t=-2e^{-2\alpha t}[1+O(e^{-4\alpha t})]$, one obtains
\begin{equation}
x^{1/2}\phi_n=\Phi_\infty(w)\big[1+O(e^{-4\alpha t})+O(x^{-2})\big]\quad\text{for}\quad xe^{-2\alpha t}=O(1),\qquad w=\ln x-2\alpha t,
\end{equation}
with $\Phi_\infty$ given in eq.~\eqref{eq:profile}, $\int\Phi_\infty^2dw=1$ and $\int e^{w}\Phi_\infty^2dw=h/2$, so that $\langle x\rangle\simeq\frac h2e^{2\alpha t}$, consistent with $K=2h\sinh^2\alpha t$.}
\begin{equation}
\Phi(t,U)\to\Phi_\infty(U-2\alpha t),\qquad
\Phi_\infty(w)=\frac{4^{h}}{\sqrt{\Gamma(2h)}}\,e^{hw-2e^{w}}, \qquad
w=\ln x-2\alpha t \,,
\label{eq:profile}
\end{equation}
with corrections $O(e^{-4\alpha t})$ in the region carrying the packet, and $\int\Phi_\infty^2dw=1$. Figure~\ref{fig:numerics} shows this for $h=\frac12$: already at $\alpha t=1$ the lattice solution is consistent with the rigid profile, and the $\ell^2$ deviation decreases as $e^{-4\alpha t}$, i.e.\ as $n^{-2}$ at the packet.

\paragraph{Chirality and the staggered sector.} The first-order equation~\eqref{eq:chiral} admits only one chirality, $\Phi=F(U-2\alpha t)$, moving toward large $n$. The opposite chirality of the second-order equation is not lost. If $\phi_n=(-1)^n\chi_n$, then $\chi_n$ obeys eq.~\eqref{eq:chain} with $t\to-t$, so smooth envelopes of staggered configurations move toward $n=0$. The second-order equation in~\eqref{eq:chiral} describes both sectors, the smooth one moving inward (toward large $n$) and the staggered one outward, consistent with the retarded/advanced structure found in the frequency domain in ref.~\cite{Dodelson:2025rng}.

%
\section{Near-horizon AdS\texorpdfstring{$_2$}{2} and the leading-order dictionary}\label{sec:leading}

\subsection{Probe scalar near the horizon}
We compare with a probe scalar field in an AdS$_2$ black hole, the geometry of JT gravity and of the near-horizon region of near-extremal black holes~\cite{Jackiw:1984je,Teitelboim:1983ux,Almheiri:2014cka,Maldacena:2016upp,Strominger:1998yg,Mertens:2022irh}:
\begin{equation}
ds^2=-f(r)\,dt^2+\frac{dr^2}{f(r)},\qquad f(r)=\frac{r^2-r_h^2}{L^2},\qquad T=\frac{r_h}{2\pi L^2} \,,
\label{eq:bh}
\end{equation}
where $r_h$ is the horizon radius, $f(r_h)=0$, and $L$ the AdS radius. We introduce the logarithmic blueshift coordinate $\zeta$,
\begin{equation}
e^{2\zeta}=\frac{L}{r-r_h},\qquad f=e^{-2\zeta}F,\qquad F=F_0+e^{-2\zeta},\qquad F_0=\frac{2r_h}{L}=4\pi TL,
\label{eq:zeta}
\end{equation}
which places the horizon at $\zeta\to\infty$. Its name reflects that the blueshift factor of static observers is $f^{-1/2}=F_0^{-1/2}e^{\zeta}[1+O(e^{-2\zeta})]$. The metric reads
\begin{equation}
ds^2=e^{-2\zeta}\Big[-F\,dt^2+\frac{4L^2}{F}\,d\zeta^2\Big],
\label{eq:metric}
\end{equation}
and near the horizon $\zeta=-2\pi T\,r_*-\frac12\ln F_0+O(e^{-2\zeta})$, where $r_*=\int dr/f$ is the tortoise coordinate (appendix~\ref{app:KG}). The Klein--Gordon equation $\Box\Phi=m^2\Phi$ reads, to first subleading order,
\begin{equation}
\pd_t^2\Phi=(2\pi T)^2\Big[\pd_\zeta^2+\frac{2}{F_0}\,e^{-2\zeta}\big(\pd_\zeta^2-\pd_\zeta-2m^2L^2\big)\Big]\Phi+O(e^{-4\zeta}),
\label{eq:KG}
\end{equation}
where $e^{-2\zeta}=(r-r_h)/L$; the exact form is given in eq.~\eqref{eq:KGexact}. At leading order every field, whatever its mass, becomes a massless wave in tortoise coordinates, and the infalling solutions are $\Phi=G(\zeta-2\pi Tt)$.

\subsection{Matching and the normalization of the radial coordinate}\label{sec:matching}
At leading order eqs.~\eqref{eq:chiral} and~\eqref{eq:KG} are both free chiral wave equations (i.e., $\Phi=F(U-2\alpha t)$ and $\Phi=G(\zeta-2\pi Tt)$), and the Krylov packets move toward large $n$ as the infalling bulk solutions move toward the horizon. Any identification of the form
\begin{equation}
U=\lambda\,\zeta+u_0,\qquad \lambda>0,
\label{eq:lambda}
\end{equation}
maps the Krylov evolution onto infalling bulk evolution, provided $2\alpha=2\pi T\lambda$, i.e.,
\begin{equation}
\alpha=\lambda\,\pi T.
\label{eq:alphalambda}
\end{equation}
The leading-order equations do not fix $\lambda$ or $u_0$. The constant $u_0$ is immaterial: it can be absorbed into the normalization of the depth, $x\to e^{u_0}x$. The ratio $\lambda$, on the other hand, is physical, since it sets $\alpha/\pi T$; it is the only invariant combination of the normalizations of the two radial coordinates. A physical input is required to fix it.

Such an input is provided by the size-momentum correspondence~\cite{Susskind:2018tei,Brown:2018kvn,Lin:2019qwu,Magan:2018nmu}: operator size, which the Krylov depth measures, is proportional to the momentum of an infalling particle relative to static observers, which grows as the blueshift factor $f^{-1/2}$. Identifying
\begin{equation}
x\propto f^{-1/2}\qquad\Longleftrightarrow\qquad U=\zeta+u_0+O(e^{-2\zeta})
\label{eq:blueshift}
\end{equation}
fixes $\lambda=1$ and hence
\begin{equation}
\alpha=\pi T .
\label{eq:dict}
\end{equation}
The same identification was obtained in ref.~\cite{Dodelson:2025rng} in the opposite direction: assuming that the effective inverse temperature $\beta_0=\pi/\alpha$ equals the inverse Hawking temperature, matching the Krylov tortoise coordinate to that of the black hole yields $n\sim(1-r_h/r)^{-1/2}\sim f^{-1/2}$. At leading order the two statements are equivalent, each implying the other. In section~\ref{sec:sl2} we show that, for the $\SL$ chain, eq.~\eqref{eq:dict} follows independently from the symmetry, without a continuum limit and without invoking eq.~\eqref{eq:blueshift}. Under eq.~\eqref{eq:blueshift} the deep-Krylov expansion parameter $x^{-2}$ is proportional to the near-horizon expansion parameter $(r-r_h)/L$.

The identification can be phrased without reference to coordinates. Along the characteristics of eq.~\eqref{eq:transport}, $dn/dt=2b(n)$, while an infalling null ray in the metric~\eqref{eq:bh}, or in any static metric $-f\,dt^2+dr^2/f$ (also in the $(t,r)$ plane of a higher-dimensional black hole), obeys $dr/dt=-f$. Demanding $n\propto f^{-1/2}$ along the ray gives
\begin{equation}
b(n)=\tfrac14\,n\,f'(r),
\label{eq:bfprime}
\end{equation}
so that near any non-extremal horizon, where $f'(r_h)=4\pi T$, $b_n\simeq\pi T\,n$. The emergent throat therefore has Hawking temperature $T_H=\alpha/\pi$. For a Krylov chain built with the Wightman inner product at temperature $T$, the bound $\alpha\le\pi T$~\cite{Parker:2018yvk,Avdoshkin:2019trj} states that $T_H\le T$: the throat seen by the chain is never hotter than the system, with equality precisely when the bound is saturated, as it is, kinematically, for conformal correlators~\cite{Dymarsky:2021bjq}. The leading-order map uses only the Rindler structure of the horizon and therefore applies to any non-extremal horizon.

\paragraph{What the leading order does not fix.}
At leading order the mass drops out of eq.~\eqref{eq:KG}, and eq.~\eqref{eq:chiral} is independent of all Lanczos data except $\alpha$. Neither the bulk mass nor any other bulk parameter can therefore be inferred from the leading-order matching. The bulk mass can be fixed exactly by representation theory in the $\SL$ case (section~\ref{sec:sl2}), but, as we show in section~\ref{sec:nogo}, it cannot be matched to local deep-Krylov data at any order.

%
\section{Exact \texorpdfstring{$\SL$}{SL(2,R)} dictionary}\label{sec:sl2}

The correspondence becomes exact when both sides carry the same $\SL$ representation.

\subsection{A quick review}
\paragraph{Bulk isometries.}
The Killing vectors of the AdS$_2$ black hole~\eqref{eq:bh} are
\begin{equation}
\xi_0=\frac{\pd_t}{2\pi T},\qquad
\xi_{\pm1}=e^{\mp2\pi Tt}\Big[\frac{r\,\pd_t}{2\pi T\sqrt{r^2-r_h^2}}\pm\sqrt{r^2-r_h^2}\,\pd_r\Big],
\label{eq:killing}
\end{equation}
which satisfy $[\xi_0,\xi_{\pm1}]=\mp\xi_{\pm1}$ and $[\xi_1,\xi_{-1}]=2\xi_0$, and whose Casimir
\begin{equation}
C=\xi_0^2-\tfrac12\big(\xi_1\xi_{-1}+\xi_{-1}\xi_1\big)=L^2\,\Box
\label{eq:casimir}
\end{equation}
is proportional to the d'Alembertian (appendix~\ref{app:sl2}). The Klein--Gordon equation is thus the Casimir equation $C\Phi=m^2L^2\Phi$, and a field of dimension $\Delta$, $m^2L^2=\Delta(\Delta-1)$, carries the lowest-weight discrete series $\Dser{\Delta}$ of single-particle states, on which the global-time generator $L_0$ has spectrum $\Delta+n$, $n=0,1,\dots$~\cite{Strominger:1998yg,Spradlin:1999bn,Kitaev:2017hnr}.

\paragraph{The Krylov chain as a discrete-series representation.}
For the coefficients~\eqref{eq:bnsl2} the Krylov states can be identified with the standard basis of $\Dser{h}$~\cite{Caputa:2021sib,Caputa:2021ori},
\begin{equation}
\ket{O_n}\equiv|h,n\rangle,\qquad L_0|h,n\rangle=(n+h)|h,n\rangle,\qquad L_{-1}|h,n\rangle=\sqrt{(n+1)(n+2h)}\,|h,n+1\rangle,
\label{eq:Dh}
\end{equation}
with $L_1=L_{-1}^\dagger$, $[L_0,L_{\pm1}]=\mp L_{\pm1}$, $[L_1,L_{-1}]=2L_0$, and the Liouvillian is
\begin{equation}
\mathcal{L}=\alpha(L_1+L_{-1})=2\alpha\,\cB,\qquad \cB\equiv\tfrac12(L_1+L_{-1}).
\label{eq:Krep}
\end{equation}
The Krylov state is the coherent state $e^{2i\alpha\cB t}|h,0\rangle=\sum_ni^n\phi_n(t)|h,n\rangle$ with $\phi_n$ given in eq.~\eqref{eq:sl2sol}.

\subsection{Temperature and mass}
The non-compact generator $i\cB$ has the same normalization as $\xi_0$: the eigenvalues of its adjoint action are $\pm1$ and $0$ (appendix~\ref{app:sl2}), and all such elements of $\mathfrak{sl}(2,\mathbb{R})$ form a single adjoint orbit. Realizing the Krylov evolution $e^{2i\alpha\cB t}$ as black-hole time evolution by the same time $t$ therefore requires
\begin{equation}
2\alpha=2\pi T\qquad\Longleftrightarrow\qquad \alpha=\pi T,
\label{eq:alphaexact}
\end{equation}
reproducing eq.~\eqref{eq:dict} without a continuum limit. Consistently, $\phi_0(t)=\cosh^{-2h}(\alpha t)$ is analytic in the strip $|{\rm Im}\,t|<\pi/2\alpha$, which coincides with the KMS strip~\cite{Kubo_1957,Martin:1959aa} $|{\rm Im}\,t|<\beta/2$ of a Wightman function if and only if $\alpha=\pi T$~\cite{Parker:2018yvk,Avdoshkin:2019trj}; in general, analyticity in the KMS strip gives only $\alpha\le\pi T$. Conversely, the Wightman two-point function of a dimension-$\Delta$ operator in the boundary theory, $\propto[\pi T/\cosh(\pi Tt)]^{2\Delta}$, has exactly $b_n=\pi T\sqrt{n(n+2\Delta-1)}$ (section~\ref{sec:sl2chain}). Hence
\begin{equation}
h=\Delta,\qquad m^2L^2=h(h-1),
\label{eq:mass}
\end{equation}
and the boundary correlators of a probe field in a fixed AdS$_2$ black hole saturate $\alpha\le\pi T$. The origin $x=n+h$ selected by the continuum analysis, eq.~\eqref{eq:originrule}, is the eigenvalue of $L_0$, and
\begin{equation}
K(t)=\langle L_0\rangle-h=h\big(\cosh 2\alpha t-1\big)=2h\sinh^2(\alpha t)
\label{eq:KL0}
\end{equation}
is the excess global energy of the boosted bulk particle.

\subsection{Breitenlohner--Freedman bound from positivity}
A Krylov chain requires real and positive $b_n^2=\alpha^2n(n+2h-1)$ for all $n\ge1$. This holds for all $n$ if and only if $h$ is real and positive, which implies
\begin{equation}
m^2L^2=h(h-1)\ \ge\ -\tfrac14 .
\label{eq:BF}
\end{equation}
A mass below the BF bound~\cite{Breitenlohner:1982bm,Breitenlohner:1982jf} corresponds to $h=\frac12\pm i\sqrt{-m^2L^2-\frac14}$, for which $b_n^2$ is complex and no Hermitian Liouvillian exists; $0<h<\frac12$ corresponds to the alternative quantization $\Delta_-$~\cite{Klebanov:1999tb}. The bound is saturated, $h=\frac12$, precisely when $b_n=\alpha n$ is exactly linear. Maximal Krylov growth, $\alpha=\pi T$, holds for every $h$; it is strict linearity, not maximal growth, that singles out the BF point. In representation-theoretic terms this is the familiar statement that unitary lowest-weight representations of $\SL$ require real $h>0$; the Krylov chain makes it concrete as positivity of the hopping amplitudes.

\subsection{The edge of the chain and quasinormal modes}
Deep in the chain $h$ enters only at relative order $n^{-2}$, eq.~\eqref{eq:Rsl2}, and, as shown in section~\ref{sec:nogo}, in a locally removable form. Its unambiguous imprint is at the edge: at fixed $n$ and late times eq.~\eqref{eq:sl2sol} gives $\phi_n(t)\propto e^{-2h\alpha t}$, which for $\alpha=\pi T$ is the lowest quasinormal mode $\omega=-2\pi iT\Delta$ of the AdS$_2$ black hole~\cite{Faulkner:2009wj,Birmingham:2001pj}. More generally, the spectral density of $\cosh^{-2\Delta}(\pi Tt)$ is $\propto|\Gamma(\Delta+i\omega/2\pi T)|^2$, whose poles in the lower half-plane, $\omega=-2\pi iT(\Delta+k)$, $k=0,1,\dots$, are the quasinormal frequencies (appendix~\ref{app:cft1}). 

In tortoise coordinates the Klein--Gordon equation reads $\pd_t^2\Phi=\pd_{r_*}^2\Phi-V\Phi$ with $V=m^2f=(2\pi T)^2m^2L^2/\sinh^2(2\pi Tr_*)$, $-\infty<r_*<0$. The potential is exponentially small in the throat, $V\propto e^{4\pi Tr_*}$, which is why the mass enters eq.~\eqref{eq:KG} only at subleading order, and grows as $m^2L^2/r_*^2$ near the boundary $r_*\to0$, where it reflects the waves and, with the infalling condition at the horizon, fixes the quasinormal frequencies. Since $x^{-2}\propto(r-r_h)/L$ under eq.~\eqref{eq:blueshift}, the edge of the chain, $n=O(1)$, where both expansions fail, is the counterpart of this region, while the deep chain is the throat, universal at leading order. This is a correspondence of regions, not a local map, which by section~\ref{sec:nogo} does not exist beyond leading order. The resulting dictionary is summarized in table~\ref{tab:dict}.

\begin{table}[t]
\centering
\renewcommand{\arraystretch}{1.15}
\begin{tabular}{@{}ll@{}}
\hline
\textbf{Krylov chain} & \textbf{Near-horizon AdS$_2$}\\
\hline
first-order Krylov evolution, smooth sector & infalling solutions\\
staggered sector $(-1)^n\chi_n$ & outgoing solutions\\
log-depth $U=\ln x$, $x=n+x_0$ & log-blueshift $\zeta=-2\pi Tr_*+{\rm const}$\\
depth $x$ & blueshift $f^{-1/2}$\\
$O(n^{-2})$ corrections & $O\big((r-r_h)/L\big)$ corrections\\
growth rate $\alpha$ & $\pi T$\\
smooth subleading $b_n$ data & no invariant content\\
\hline
$L_0=n+h$ & global AdS$_2$ energy\\
$\mathcal{L}=2\alpha\cB$ & $i\pd_t=2\pi T\,(i\xi_0)$\\
weight $h$ (Casimir $h(h-1)$) & $\Delta$, $m^2L^2=\Delta(\Delta-1)$\\
$K=\langle L_0\rangle-h$ & excess global energy\\
$b_n^2>0$ & BF bound $m^2L^2\ge-\frac14$\\
$b_n=\alpha n$ exactly ($h=\frac12$) & BF saturation\\
edge decay $e^{-2h\alpha t}$ & lowest quasinormal mode $-2\pi iT\Delta$\\
\hline
\end{tabular}
\caption{Dictionary between the Krylov chain and the AdS$_2$ black hole. The upper block holds at leading order in $n^{-2}$ for any $b_n\simeq\alpha n$ (the identification $x\propto f^{-1/2}$ is the input that fixes $\alpha=\pi T$; the last row of the block is proven in section~\ref{sec:nogo}). The lower block is exact for $b_n=\alpha\sqrt{n(n+2h-1)}$.}
\label{tab:dict}
\end{table}

%
\section{Beyond leading order: no local matching of the mass}\label{sec:nogo}

\subsection{Subleading Lanczos data enter at the same order}
\label{sec:sameorder}

The leading near-horizon corrections in eq.~\eqref{eq:KG} are of relative order $e^{-2\zeta}=(r-r_h)/L$. Under the identification~\eqref{eq:blueshift} they are of the same order as the $x^{-2}$ corrections in eq.~\eqref{eq:subleading}, which receive two contributions: the higher-derivative terms of the lattice operator and the subleading Lanczos coefficient $\beta_1$. In particular, the information beyond Rindler carried by eq.~\eqref{eq:KG} cannot be compared with the leading linear growth alone; the subleading data of $b_n$ must be included at the same order, as emphasized in ref.~\cite{Dodelson:2025rng}. Both effects are contained in $R(\cD)$. The question is whether, at this order and beyond, the two equations can be identified, and for which masses.

Since the derivative order can be changed by field redefinitions, the comparison must be made modulo transformations that do not change the physical content. We write both second-order equations in the form
\begin{equation}
\pd_t^2\Phi=\alpha^2\big[4\cD^2+e^{-2U}V(\cD)\big]\Phi+O(e^{-3U}),
\label{eq:Vform}
\end{equation}
with $U$ the Krylov log-depth on one side and $U=\zeta$ on the other, so that $e^{-2U}=e^{-2\zeta}$ [the identification~\eqref{eq:blueshift} with $u_0=0$; a nonzero $u_0$ rescales $V$ by the positive factor $e^{2u_0}$], and $\alpha=\pi T$. From eqs.~\eqref{eq:subleading} and~\eqref{eq:KG} (appendix~\ref{app:logdepth}),
\begin{equation}
V_{\rm K}(\cD)=4(\cD-1)\,R(\cD),\qquad
V_{\rm G}(\cD)=\frac{8}{F_0}\big(\cD^2-\cD-2m^2L^2\big),
\label{eq:VKVG}
\end{equation}
where the subscripts K and G refer to the Krylov and gravity sides, and $R(\cD)$ is given in eq.~\eqref{eq:subleading}.

\subsection{Allowed transformations and the invariant}
\label{sec:invariant}

We compare the equations up to transformations that are static (the Krylov time is identified with the Killing time, as fixed at leading order), local, and analytic in $z\equiv e^{-2U}$ near $z=0$:
\begin{enumerate}
\item[(a)] field redefinitions $\Phi=[1+zQ(\cD,\pd_t)]\tilde\Phi$ with $Q$ polynomial;
\item[(b)] changes of radial coordinate $U\to U+\lambda_1 z+O(z^2)$;
\item[(c)] changes of time slicing $t\to t+G(U)$ with $G$ analytic in $z$;
\item[(d)] multiplication of the equation by an invertible local operator $1+zM$,
\end{enumerate}
together with the use of the leading-order equation $\pd_t^2=4\alpha^2\cD^2$ inside $O(z)$ terms. Using $\cD^2e^{-2U}=e^{-2U}(\cD-2)^2$, (a) shifts $V\to V-16(\cD-1)Q$, (b) shifts $V\to V+16\lambda_1\cD(\cD-1)$, and (d) leaves $V$ unchanged (appendix~\ref{app:transf}).

The cleanest formulation is in terms of solutions. For $\Phi\propto e^{-i\omega t}$ and
\begin{equation}
k\equiv\frac{\omega}{2\alpha},
\label{eq:k}
\end{equation}
the solution of eq.~\eqref{eq:Vform} with leading behavior $e^{ikU}$, the ingoing one, is
\begin{equation}
\Phi=e^{-i\omega t}\,e^{ikU}\big[1+c(k)\,e^{-2U}+O(e^{-3U})\big],\qquad c(k)=\frac{V(ik)}{16\,(ik-1)} .
\label{eq:ck}
\end{equation}
The denominator vanishes at $ik=1$, i.e., $\omega=-2i\alpha$, where the ingoing exponent $ik-2$ of the subleading term coincides with the outgoing exponent $-ik$. For $\alpha=\pi T$ this is the first special point $\omega_1=-2\pi iT$ of ingoing near-horizon solutions~\cite{Blake:2019otz,Blake:2017ris,Grozdanov:2017ajz,Blake:2018leo,Natsuume:2019xcy}. Transformations (a)--(d) change $c(k)$ only by entire functions of $k$, polynomials for (a), (b), (d) and, for (c), the expansion coefficients of the entire factor $e^{-i\omega G(U)}$; hence the residue of $c(k)$ at $ik=1$ is invariant. Conversely, if $V(1)=0$, then $V=(\cD-1)W$ with $W$ polynomial and the redefinition (a) with $Q=W/16$ removes $V$ entirely. Hence $V(1)$, up to a positive factor, is the complete invariant at this order. From eq.~\eqref{eq:VKVG},
\begin{equation}
V_{\rm K}(1)=0\quad\text{for every }\beta_1\text{ (in particular every }h\text{)},\qquad V_{\rm G}(1)=-\frac{16\,m^2L^2}{F_0}.
\label{eq:V1}
\end{equation}
At $O(n^{-2})$ the deep-Krylov equation is therefore locally equivalent to a massless field, independently of $\beta_1$, while the near-horizon Klein--Gordon equation is not, unless $m=0$. For $m=0$ the two are related by the field redefinition $\Phi_{\rm K}=[1+e^{-2U}Q(\cD)]\Phi_{\rm G}$ with $Q=(V_{\rm K}-V_{\rm G})/[16(\cD-1)]$.

\subsection{All orders: smooth chains are pole-free}
\label{sec:allorders}

The vanishing of $V_{\rm K}(1)$ is not an accident of the first correction. It reflects a structural property of the Krylov equation: it is first order in time.\footnote{The author acknowledges the use of Claude Opus 5.5 (Anthropic) in formulating and checking Proposition~\ref{prop:poly}, its proof and Corollary~\ref{cor:second} in section~\ref{sec:nogo}.}

\begin{proposition}
\label{prop:poly}
Let the Lanczos coefficients admit an asymptotic expansion $b_n\sim\alpha n+\sum_{j\ge0}a_jn^{-j}$ with $\alpha>0$, and let $x=n+x_0$ for any $x_0$. For every $\omega\in\mathbb{C}$ the Krylov equation $-i\omega\phi_n=b_n\phi_{n-1}-b_{n+1}\phi_{n+1}$ has a unique formal solution of the form
\begin{equation}
\phi_n=x^{ik-\frac12}\sum_{Q\ge0}c_Q(k)\,x^{-Q},\qquad c_0=1,\qquad k=\frac{\omega}{2\alpha},
\label{eq:formal}
\end{equation}
and every coefficient $c_Q(k)$ is a polynomial in $k$. In particular, all coefficients are entire functions of the frequency.
\end{proposition}

\begin{proof}
Write $b_n=B(x-\frac12)$ and $b_{n+1}=B(x+\frac12)$ with $B(y)=\alpha y+\tilde B(y)$, where $\tilde B(y)\sim\beta_0+\sum_{j\ge1}\beta_jy^{-j}$, and substitute $\phi(x)=\sum_qc_qx^{s-q}$, $s=ik-\frac12$, into
\begin{equation}
E\equiv B(x-\tfrac12)\,\phi(x-1)-B(x+\tfrac12)\,\phi(x+1)+i\omega\,\phi(x)=0 .
\end{equation}
The terms linear in $\alpha$ give, for each $q$,
\begin{equation}
\begin{aligned}
&\alpha x\big[(x-1)^{s-q}-(x+1)^{s-q}\big]-\tfrac\alpha2\big[(x-1)^{s-q}+(x+1)^{s-q}\big]+i\omega x^{s-q}\\
&\qquad=\big[2\alpha q-2\alpha(s+\tfrac12)+i\omega\big]x^{s-q}+O(x^{s-q-2})=2\alpha q\,x^{s-q}+O(x^{s-q-2}),
\end{aligned}
\end{equation}
since $2\alpha(s+\frac12)=2i\alpha k=i\omega$. The terms involving $\tilde B$ are $O(x^{s-q-1})$. All coefficients in these expansions are polynomials in $s$, hence in $k$, because they arise from the binomial series of $(x\mp1)^{s-q}$ and from the expansion of $\tilde B(x\mp\frac12)$ in $1/x$. Collecting the power $x^{s-Q}$ gives
\begin{equation}
2\alpha Q\,c_Q=-\sum_{q<Q}P_{Q,q}(k)\,c_q,
\label{eq:recursion}
\end{equation}
with polynomials $P_{Q,q}$. At the leading power $x^{s}$ the coefficient of $c_0$ is $2\alpha\cdot0=0$; this is the indicial equation, which fixes $s=ik-\frac12$. Since the coefficient $2\alpha Q$ of $c_Q$ is independent of $k$ and nonzero for $Q\ge1$, induction on $Q$ shows that every $c_Q$ is a polynomial in $k$.
\end{proof}

\begin{corollary}
\label{cor:second}
The ingoing formal solution of the second-order equation $\pd_t^2\phi=\Lk^2\phi$ coincides with eq.~\eqref{eq:formal}. Its recursion divides by $4\alpha^2Q(Q-2ik)$, which vanishes at $ik=Q/2$, but all these apparent poles are removable. In particular, the ingoing near-horizon solutions of any smooth Krylov chain have no singularity at any special point $\omega=-iQ\alpha$, at any order in $1/n$.
\end{corollary}

\begin{proof}
A solution of the first-order equation solves the second-order one. Away from $ik=Q/2$ the ingoing formal solution of the second-order equation is unique, so it coincides with eq.~\eqref{eq:formal}; its coefficients are polynomials there and hence everywhere.
\end{proof}

At the first nontrivial order, $c_1=0$ when the origin is chosen by eq.~\eqref{eq:originrule}, and $c_2=R(ik)/4$, which agrees with eq.~\eqref{eq:ck} for $V_{\rm K}=4(\cD-1)R$. If the first correction to the leading equation appears at order $e^{-pU}$, as $\pd_t\Phi=-\alpha[2\cD+e^{-pU}R_p(\cD)]\Phi$, the second-order potential is $V_p=2(2\cD-p)R_p$, which vanishes precisely at the corresponding resonance $\cD=p/2$.

\subsection{The bulk: a pole at the first special point}
\label{sec:bulkpole}

On the gravity side, the near-horizon expansion of the ingoing solution follows from the exact solution of the radial equation (appendix~\ref{app:exactKG}). With $X=r/r_h$ and $w=\omega/2\pi T$ (so that $w=k$ for $\alpha=\pi T$), the radial equation is the associated Legendre equation, and the ingoing solution is
\begin{equation}
R_{\rm in}=\Big(\frac{X+1}{X-1}\Big)^{iw/2}{}_2F_1\Big(1-\Delta,\Delta;1-iw;\frac{1-X}{2}\Big)
\propto(X-1)^{-iw/2}\big[1+a_1(X-1)+\dots\big],
\label{eq:Rin}
\end{equation}
with $\Delta(\Delta-1)=m^2L^2$ and
\begin{equation}
a_1=\frac{w^2+iw+2m^2L^2}{4(1-iw)} .
\label{eq:a1}
\end{equation}
The coefficient $a_1$ is exactly $c(k)\,r_h/L$ obtained from $V_{\rm G}$, and it has a pole at $w=-i$, $\omega=-2\pi iT$, with
\begin{equation}
\lim_{w\to-i}(1-iw)\,a_1=\frac{m^2L^2}{2}.
\label{eq:residue}
\end{equation}
More generally, the expansion coefficients of the hypergeometric function have poles at $1-iw=0,-1,\dots$, i.e., at the special points $\omega_n=-2\pi iTn$, with residues proportional to products of $(1-\Delta)_j(\Delta)_j$, the first being $\propto(1-\Delta)\Delta=-m^2L^2$. For $m=0$ ($\Delta=1$) the hypergeometric function equals $1$ and the near-horizon expansion is pole-free to all orders, as it must be: in two dimensions the massless Klein--Gordon equation is conformally invariant and reduces exactly to the flat wave equation in $(t,r_*)$, so the curvature of the throat is invisible to it.

\subsection{Consequences}
\label{sec:consequencesnogo}

Combining sections~\ref{sec:invariant}--\ref{sec:bulkpole}: \emph{no static, local, analytic identification of the deep region of a smooth Krylov chain with the near-horizon region of an AdS$_2$ black hole exists beyond leading order for a massive probe field.} The obstruction appears at the first subleading order and is independent of the subleading Lanczos data. For $m=0$, on the other hand, both sides have ingoing solutions with coefficients polynomial in the frequency, and they are related by a local field redefinition order by order (appendix~\ref{app:massless}). In particular, $h$ and $m^2$ cannot be related by local near-horizon matching; the relation $m^2L^2=h(h-1)$ of eq.~\eqref{eq:mass} is global. The mass is encoded at the edge of the chain, in the late-time decay governed by the lowest quasinormal mode, consistent with the fact that in the bulk the mass controls the reflection of waves near the AdS$_2$ boundary.

This result sharpens the observation of ref.~\cite{Dodelson:2025rng} that the leading linear growth of $b_n$ encodes only the existence of a horizon, while information beyond Rindler requires subleading data. For a probe scalar in AdS$_2$, the only such information at the first subleading order is the residue~\eqref{eq:residue}, and smooth subleading data of any order cannot supply it. It also extends to all orders the first-order observation of ref.~\cite{Dodelson:2025rng} that smooth corrections to $b_n$ produce entire functions of the frequency.

Within the Krylov chain, a local mechanism that does produce the required singularity is a non-smooth component of the Lanczos coefficients. For $b_n=\bar b_n+(-1)^n\varepsilon\,\alpha\,x^{-\gamma}$ with $\bar b_n$ smooth, the smooth ingoing component sources the staggered (outgoing) sector, with amplitude (appendix~\ref{app:staggered})
\begin{equation}
\phi_n\supset(-1)^n\,\sigma\,x^{ik-\frac12-\gamma},\qquad \sigma=\frac{\varepsilon}{\gamma-2ik}+O(\varepsilon^2),
\label{eq:sigma}
\end{equation}
which has a pole at $\omega=-i\gamma\alpha$, reproducing in the time-domain language the pole found in ref.~\cite{Dodelson:2025rng} for staggered corrections. The resonance occurs because $x^{ik-\frac12-\gamma}$ coincides, at $ik=\gamma/2$, with the free outgoing exponent $x^{-ik-\frac12}$ of the staggered sector, the same mechanism as in the bulk. For $\gamma=2$, a staggered component of relative order $n^{-3}$ in $b_n$, the pole sits exactly at the first special point $\omega_1=-2\pi iT$, with residue $\propto\varepsilon$. A staggered $n^{-2}$ term is thus the simplest local Krylov data with the analytic structure of a bulk mass; staggered Lanczos sequences do occur, for instance in mass-deformed SYK~\cite{Dodelson:2025rng} and in theories with a gapped or bounded spectrum~\cite{Camargo:2022rnt}. In section~\ref{sec:beyond} we show, however, that the BTZ Lanczos sequence contains no such component, and argue that the same holds for holographic Lanczos sequences in general: the near-horizon data are encoded non-locally, in the spectral factorization of the Wightman density. The $\SL$ chain is the simplest example: it has no staggered component, and the bulk mass is encoded through the exact $\SL$ intertwiner between Krylov states and bulk single-particle states.

%
\section{Beyond AdS\texorpdfstring{$_2$}{2}: BTZ, black branes and the spectral factor}\label{sec:beyond}

Section~\ref{sec:nogo} shows that a smooth Krylov chain cannot reproduce locally the near-horizon singularity carried by a massive field in AdS$_2$. Two questions remain. Does the obstruction persist beyond AdS$_2$, where the near-horizon potential receives contributions from momentum and from the transverse geometry? And do the Lanczos sequences of actual holographic correlators contain non-smooth components, such as the staggered terms of eq.~\eqref{eq:sigma}, that could supply the singularity? We answer both questions, analytically for the BTZ black hole~\cite{Banados:1992wn} at finite momentum, and identify the mechanism that makes holographic Lanczos sequences smooth.

\subsection{BTZ at finite momentum: exact Lanczos coefficients}
\label{sec:btz}

The planar BTZ black hole is dual to a two-dimensional CFT on a line at temperature $T$. For a scalar primary of weights $h=\bar h=\Delta/2$, the half-shifted (Wightman) correlator is $\propto[\pi T/\cosh\pi T(x-t)]^{\Delta}[\pi T/\cosh\pi T(x+t)]^{\Delta}$. Its Fourier transform at momentum $q$ factorizes in light-cone coordinates (appendix~\ref{app:btzft}), giving the Wightman spectral density of the mode $O_q$,
\begin{equation}
\rho_q(\omega)\propto\Big|\Gamma\Big(\frac\Delta2+i\frac{\omega+q}{4\pi T}\Big)\Big|^2\Big|\Gamma\Big(\frac\Delta2+i\frac{\omega-q}{4\pi T}\Big)\Big|^2 .
\label{eq:rhobtz}
\end{equation}
In the variable $x=\omega/4\pi T$ this is the weight $\Gamma(a+ix)\Gamma(b+ix)\Gamma(c-ix)\Gamma(d-ix)$ of the continuous Hahn polynomials~\cite{Koekoek_2010}, with $a=\bar c=\frac\Delta2+\frac{iq}{4\pi T}$ and $b=\bar d=\frac\Delta2-\frac{iq}{4\pi T}$. Their three-term recurrence (appendix~\ref{app:btzhahn}) gives vanishing diagonal coefficients and
\begin{equation}
b_n=2\pi T\,\sqrt{\frac{\big(m^2-(\Delta-1)^2\big)\big(m^2+Q^2\big)}{4m^2-1}},\qquad m=n+\Delta-1,\qquad Q=\frac{q}{2\pi T}.
\label{eq:bnbtz}
\end{equation}
For instance, $\Delta=1$ and $q=0$ give $b_n=2\pi T\,n^2/\sqrt{4n^2-1}$. To our knowledge this closed form has not appeared in the Krylov literature; the Lanczos coefficients of two-dimensional CFTs have so far been studied mostly at fixed spatial position or within the $\SL$ family.

Three properties are relevant here.
\begin{enumerate}
\item $\alpha=\pi T$ for every $\Delta$ and $q$: the throat seen by the chain is the Rindler region of the BTZ horizon, in agreement with section~\ref{sec:matching}.
\item $b_n$ is the square root of a rational function of $n$, so it has a smooth large-$n$ expansion and \emph{no staggered component at any order}.
\item The expansion
\begin{equation}
b_n=\pi T\Big[m+\frac{Q^2-(\Delta-1)^2+\frac14}{2m}+O(m^{-3})\Big]
\end{equation}
gives the origin $x_0=\Delta-\frac12$ of eq.~\eqref{eq:originrule} and $\beta_1=\frac{\pi T}{2}\big(Q^2-(\Delta-1)^2+\frac14\big)$. The momentum enters the deep chain only through $\beta_1$, which by Proposition~\ref{prop:poly} carries no invariant local content. The effective Casimir of section~\ref{sec:origin}, $h_{\rm eff}(h_{\rm eff}-1)=(\Delta-1)^2-Q^2-\frac12$, falls below $-\frac14$ whenever $Q^2>(\Delta-1)^2-\frac14$ (for example for $\Delta=1$ at any $q$), although the chain is perfectly regular: the local $O(n^{-2})$ data have no representation-theoretic meaning.
\end{enumerate}

\subsection{BTZ in the bulk: the special-point residue}
\label{sec:btzbulk}

In the bulk, with $\rho=r/r_h$, $w=\omega/2\pi T$ and $m^2=\Delta(\Delta-2)$ (in units of the AdS$_3$ radius), the radial equation of $\Phi=e^{-i\omega t+iqx}R(\rho)$ is
\begin{equation}
\frac1\rho\big(\rho(\rho^2-1)R'\big)'+\Big[\frac{w^2}{\rho^2-1}-\frac{Q^2}{\rho^2}-m^2\Big]R=0 .
\label{eq:btzradial}
\end{equation}
The ingoing solution $R=\varepsilon^{-iw/2}[1+a_1\varepsilon+\dots]$, $\varepsilon=\rho-1$, has (appendix~\ref{app:btzbulk})
\begin{equation}
a_1=\frac{w^2+3iw+2Q^2+2m^2}{4(1-iw)},\qquad \lim_{w\to-i}(1-iw)\,a_1=\frac{Q^2+(\Delta-1)^2}{2}.
\label{eq:btzres}
\end{equation}
The residue vanishes only at $q=\pm2\pi iT(\Delta-1)$, which are exactly the pole-skipping points of the BTZ retarded Green's function at $\omega=-2\pi iT$~\cite{Blake:2019otz,Son:2002sd,Natsuume:2020snz}. Since the invariance argument of section~\ref{sec:invariant} is formulated in terms of solutions, and the factor relating $R$ to the field of the tortoise-coordinate problem is entire in $\omega$, the conclusion of section~\ref{sec:nogo} carries over: \emph{for the BTZ black hole the deep Krylov chain admits no static, local, analytic identification with the near-horizon region beyond leading order, for every real momentum and every $\Delta$ except $\Delta=1$, $q=0$.} In contrast with AdS$_2$, the obstruction is present also for a massless bulk field ($\Delta=2$), with residue $(Q^2+1)/2$. It is therefore not an artefact of the conformal invariance of the massless two-dimensional wave equation.

\subsection{Why holographic Lanczos sequences are smooth}
\label{sec:factor}

The absence of staggering in eq.~\eqref{eq:bnbtz} has a general origin. Consider a static black brane with a non-degenerate horizon, and let $\phi_{\rm in}$ be the ingoing radial solution, in a radial coordinate $u$ with the boundary at $u=0$ and the horizon at $u=1$, normalized at the horizon as $(1-u)^{-iw/2}[1+O(1-u)]$, with $w=\omega/2\pi T$. Denote by $A(w)$ the coefficient of its non-normalizable (source) mode at the boundary. Conservation of the radial flux gives ${\rm Im}\,G_R(\omega)\propto w/|A(w)|^2$, and the Wightman density is ${\rm Im}\,G_R/\sinh(\beta\omega/2)$. Continuing from the real axis,
\begin{equation}
\rho_W(w)\propto\frac{w}{A(w)\,A(-w)\,\sinh\pi w}.
\label{eq:factor}
\end{equation}
The function $A(w)$ is meromorphic. Its zeros are the quasinormal frequencies, and its only poles are at the special points $w=-in$, $n=1,2,\dots$, where the Frobenius recursion for $\phi_{\rm in}$ breaks down. There the pole is simple, and its residue is proportional to the near-horizon residue of section~\ref{sec:bulkpole} (times the source coefficient of the solution with the other horizon exponent). Correspondingly, $A(-w)$ has poles at $w=+in$, while $\sinh\pi w$ has simple zeros at all $w=in$, $n\in\mathbb{Z}$. Three consequences follow.
\begin{enumerate}
\item \emph{$\rho_W$ has no zeros in the complex plane.} The special-point poles of the horizon-normalized solution are cancelled by the zeros of the thermal factor, and the zero of $\sinh\pi w$ at $w=0$ is cancelled by the numerator. This is the property that ref.~\cite{Dodelson:2025rng} noted for holographic theories and related to the smoothness of the Lanczos coefficients; eq.~\eqref{eq:factor} traces it to the horizon.
\item The poles of $\rho_W$ are the quasinormal frequencies and their mirror images, together with those Matsubara frequencies $\omega=\pm2\pi iTn$ at which the residue of $A$ vanishes. This happens when the near-horizon residue vanishes, as at pole-skipping points, or when the solution with the other horizon exponent is normalizable at the boundary. For AdS$_2$ with $m=0$ ($\Delta=1$) the former occurs at every $n$, and indeed $|\Gamma(1+iw)|^2$ has poles at all $w=\pm in$.
\item The near-horizon residues are therefore not visible as local features of the spectral density, and hence of the Lanczos sequence. They are encoded in the spectral factor $A$ of $w/(\rho_W\sinh\pi w)$, the factor analytic and zero-free in the upper half-plane (unique up to the usual growth conditions), which is a global functional of the whole sequence $\{b_n\}$.
\end{enumerate}
The two exact cases make this explicit:
\begin{equation}
A_{\rm AdS_2}\propto\frac{\Gamma(1-iw)}{\Gamma(\Delta-iw)},\qquad
A_{\rm BTZ}\propto\frac{\Gamma(1-iw)}{\Gamma\big(\frac\Delta2-\frac{i(w+Q)}2\big)\,\Gamma\big(\frac\Delta2-\frac{i(w-Q)}2\big)} .
\label{eq:Aexact}
\end{equation}
Inserted in eq.~\eqref{eq:factor}, these reproduce $|\Gamma(\Delta+iw)|^2$ and eq.~\eqref{eq:rhobtz}. The factor $\Gamma(1-iw)$ supplies the special-point poles, and the residue of $A_{\rm BTZ}$ at $w=-i$, $\propto1/[\Gamma(\frac{\Delta-1-iQ}{2})\Gamma(\frac{\Delta-1+iQ}{2})]$, vanishes at the pole-skipping momenta $Q=\pm i(\Delta-1)$, where eq.~\eqref{eq:btzres} vanishes. Otherwise it vanishes only at $Q=\pm i(\Delta+1+2k)$, $k\ge0$, where the solution with the other horizon exponent is normalizable.

%
\section{Power-law Lanczos coefficients}\label{sec:power}

For $b_n=\alpha n^p$ with $p\neq1$ the system departs from exponential growth~\cite{Parker:2018yvk}. The leading-order equation~\eqref{eq:transport} still describes rigid transport, now in the Krylov tortoise coordinate $s(x)=x^{1-p}/[2\alpha(1-p)]$. For a packet of width $\Delta s=O(1)$, i.e., $\Delta x\sim x^p$, the corrections in eq.~\eqref{eq:Lexp} are suppressed by $x^{-2p}$ for $0<p\le1$ and by $x^{-2}$ for $p>1$, so the description is controlled for $p>0$ but not for $p=0$. In the variables~\eqref{eq:logdepth}, $2B\pd_x+B'=2\alpha x^{p-1}[\cD+\frac{p-1}{2}]$ on $\Phi$, and the field $\Psi\equiv x^{p-1}\Phi$ obeys
\begin{equation}
\pd_t^2\Psi=4\alpha^2\,e^{2(p-1)U}\Big[\pd_U^2-\frac{(p-1)^2}{4}\Big]\Psi
\label{eq:pwave}
\end{equation}
up to subleading corrections. The $U$-dependent prefactor does not signal a breakdown of Lorentz invariance: in $1+1$ dimensions a position-dependent propagation speed is equivalent to a conformally flat metric $-dt^2+c(U)^{-2}dU^2$, and in the coordinate $s$ the equation is again the free chiral wave equation. Along characteristics, $dn/dt=2\alpha n^p$, so
\begin{equation}
n(t)\simeq
\begin{cases}
[2\alpha(1-p)\,t]^{1/(1-p)}, & p<1,\\
e^{2\alpha t}, & p=1,\\
[2\alpha(p-1)(t_c-t)]^{-1/(p-1)}, & p>1.
\end{cases}
\label{eq:front}
\end{equation}
For $0<p\le1$ the Krylov complexity, the mean of $n$ over the rigidly translated packet, follows the same law at late times, with $K/n(t)\to1$ for $p<1$; for $b_n=\alpha\sqrt n$ one has exactly $\phi_n=e^{-\alpha^2t^2/2}(\alpha t)^n/\sqrt{n!}$ and $K=(\alpha t)^2$. For $p=0$ only the exponent survives: for constant $b_n=b$, $K\simeq\frac{16}{3\pi}\,bt$.

The invariant content is the relation between depth and tortoise coordinate. Applying eq.~\eqref{eq:bfprime} with $n\propto f^{-1/2}$, a power law $b\propto n^p$ gives $f'\propto f^{(1-p)/2}$, hence
\begin{equation}
f\propto(r-r_h)^{2/(1+p)}
\label{eq:fpower}
\end{equation}
near $r_h$. For $p\ge0$, only $p=1$ (simple zero: non-extremal, $T=\alpha/\pi$) and $p=0$ (double zero: extremal, $T=0$) give regular horizons; the latter is a kinematic statement, since the continuum description is not controlled at $p=0$. For other $p\ge0$ the curvature $\mathcal{R}=-f''$ diverges at $r=r_h$. For $p>1$, $\int dr/f<\infty$: the horizon lies at finite tortoise distance. Its Krylov counterpart is $\sum_n1/b_n<\infty$, for which the wavefront~\eqref{eq:front} reaches $n=\infty$ at finite $t_c$.

%
\section{Discussion}\label{sec:discussion}
We have established a dictionary between the dynamics of the Krylov chain and near-horizon AdS$_2$ physics at two levels, and determined where it stops being local.

At leading order in $n^{-2}$, valid for any asymptotically linear $b_n$, the unitary continuum limit of the Krylov equation is chiral transport in $\ln n$, which coincides with infalling waves in the AdS$_2$ throat; the first-order Krylov evolution supplies the infalling boundary condition, and the staggered sector supplies the outgoing one. The matching fixes only the relative normalization $\lambda$ of the two radial coordinates; the depth-blueshift identification, equivalent at this order to the one of ref.~\cite{Dodelson:2025rng}, sets $\lambda=1$ and $\alpha=\pi T$, and takes the general form $b(n)=\frac14nf'(r)$. Since this map uses only the Rindler structure of the horizon, it applies to any non-extremal horizon, including the AdS$_2$ throats of near-extremal black holes in higher dimensions.

For $\SL$ Lanczos coefficients the map is exact. The Krylov chain is the bulk single-particle representation $\Dser{\Delta}$, the Liouvillian is $2\pi T$ times the boost generator, $\alpha=\pi T$ follows from the symmetry, Krylov complexity is the excess global energy, and positivity of $b_n^2$ implies the BF bound, saturated by strictly linear growth.

Beyond leading order, subleading Lanczos data enter at the same order as the near-horizon corrections of the Klein--Gordon equation. We have shown that, for smooth Lanczos sequences, these data have no invariant content under local field redefinitions: the ingoing solutions are polynomial in frequency to all orders, whereas a massive bulk field has a pole at the first special point with residue $\propto m^2$. The AdS$_2$-specific information, $h=\Delta$, is therefore not local in the deep chain; it is encoded at the edge, in agreement with the role of the mass in reflecting waves near the AdS boundary. The same holds beyond AdS$_2$. For BTZ at finite momentum, the Lanczos coefficients are known in closed form and are smooth, and the bulk residues, which depend on mass and momentum, are generically non-zero. The mechanism is general: the thermal factor of the Wightman density cancels the special-point poles of the horizon-normalized bulk solution, so the density has no zeros and the near-horizon residues sit in its spectral factor. A complete dictionary beyond leading order is therefore necessarily non-local in $n$, as is the exact $\SL$ intertwiner.

Relative to the frequency-domain scattering picture of ref.~\cite{Dodelson:2025rng}, our results provide a manifestly unitary time-domain formulation with controlled corrections, an exact $\SL$ completion, an all-orders statement on the analytic structure of smooth chains, a second exactly solvable holographic chain (BTZ at finite momentum), and an explicit separation between universal near-horizon data ($\alpha$) and non-universal data ($h$, $q$) carried by the edge and the spectral factor. They complement holographic Krylov complexity~\cite{Jian:2020qpp,Lin:2022rbf,Rabinovici:2023yex,Heller:2024ldz,Ambrosini:2024sre,Fu:2025kkh,Caputa:2021sib}, which tracks the global growth of the Einstein-Rosen bridge, by describing the evolution of the Krylov state itself.

The dictionary yields quantitative predictions that can be tested with exact diagonalization or tensor-network computations of Lanczos coefficients in SYK-like models and chaotic spin chains: (1) in the deep Krylov region the profile of $\phi_n(t)$ translates rigidly in $\ln(n+x_0)$ at speed $2\alpha$, with deviations $O(n^{-2})$ only when the origin is fixed by the offset of $b_n$, eq.~\eqref{eq:originrule}, and $O(n^{-1})$ otherwise; (2) at $O(n^{-2})$ the local dynamics depends on the Lanczos data only through $\beta_1/\alpha$; (3) amplitudes at fixed $n$ decay at the rate set by the leading singularity of the spectral function, the lowest quasinormal frequency in holographic systems; (4) holographic Lanczos sequences have no staggered component, and their offset follows $b_n\simeq\pi T(n+\Delta-\frac d2)$, observed here for $d=1,2$.

Several directions remain open: an explicit construction of the non-local map between the deep chain and the throat beyond leading order; the Krylov counterpart of the higher special points and of pole-skipping in more general geometries~\cite{Blake:2019otz,Blake:2017ris,Grozdanov:2017ajz,Blake:2018leo}; a proof that zero-free spectral densities yield smooth Lanczos sequences at all orders; an $\SL\times\SL$ interpretation of the BTZ chain, suggested by the appearance of continuous Hahn functions as Clebsch-Gordan coefficients of $\SL$~\cite{Groenevelt:aa}; the extension to double-scaled SYK and sine-dilaton gravity, where Krylov complexity is known to compute bulk lengths~\cite{Heller:2024ldz,Ambrosini:2024sre,Fu:2025kkh}; and the inclusion of $1/N$ corrections, which should characterize the transition from the throat to the asymptotic region.

%
\acknowledgments
HSJ is supported by an appointment to the JRG Program at the APCTP through the Science and Technology Promotion Fund and Lottery Fund of the Korean Government. HSJ is also supported by the Korean Local Governments -- Gyeongsangbuk-do Province and Pohang City.
During the preparation of this manuscript, the author used Claude Opus 5.5 (Anthropic) to draft the text based on the author's own calculations, results, and manuscript outline. The author thoroughly reviewed and edited the final text, including all equations and references, and takes full responsibility for the content of the publication.

%
\appendix

%
\section{Derivative expansion of the Krylov generator}
\label{app:expansion}
\subsection{General expansion}
With $b_n=B(x_n-\frac12)$ and $b_{n+1}=B(x_n+\frac12)$, the Taylor expansion of eq.~\eqref{eq:Lexact} gives
\begin{equation}
\Lk\phi=\sum_{j,k\ge0}\frac{(\frac12)^j\big[(-1)^{j+k}-1\big]}{j!\,k!}\,B^{(j)}\phi^{(k)}
=-2\!\!\sum_{j+k\ \mathrm{odd}}\frac{B^{(j)}\phi^{(k)}}{2^j\,j!\,k!},
\label{eq:series}
\end{equation}
which yields eq.~\eqref{eq:Lexp} at total orders $j+k=1,3$. The adjoint on $L^2(dx)$ preserves the total order $j+k$, and $\Lk^\dagger=-\Lk$ holds identically in $B$; hence each homogeneous order is skew-adjoint by itself. For $B=\alpha x$, $\Lk=\alpha[(x-\tfrac12)e^{-\pd}-(x+\tfrac12)e^{\pd}]=-\alpha(2x\sinh\pd+\cosh\pd)=-\alpha\{x,\sinh\pd\}$.

\subsection{Inconsistent truncation}
\label{app:truncation}
In the variable $\tilde x=n$ (so that $1+2\tilde x=2x$ for $b_n=\alpha n$), a Taylor expansion of eq.~\eqref{eq:chain} truncated at second derivatives gives $\pd_t\phi=-\alpha[1+(1+2\tilde x)\pd+\frac12\pd^2]\phi$, i.e., the first bracket of eq.~\eqref{eq:Lexp} plus only the term $-\frac{\alpha}{2}\pd^2$ of the second. The omitted term $-\frac{\alpha}{3}x\pd^3$ is of the same order, $O(x^{-2})$, as the retained one, and only their sum is skew-adjoint. The truncated operator is not: for decaying $\phi$,
\begin{equation}
\frac{d}{dt}\int\phi^2dx=-2\alpha\int\phi\Big[\phi+2x\phi'+\tfrac12\phi''\Big]dx=\alpha\int(\phi')^2dx>0,
\end{equation}
and its Fourier modes grow as $e^{\alpha k^2t/2}$ (backward diffusion), so the initial-value problem is ill-posed. Squaring this operator and truncating again yields $\pd_t^2\phi=\alpha^2[4(1+\tilde x+\tilde x^2)\pd^2+4(1+2\tilde x)\pd+1]\phi$, whereas the exact lattice relation
\begin{equation}
\pd_t^2\phi_n=b_nb_{n-1}\phi_{n-2}-(b_n^2+b_{n+1}^2)\phi_n+b_{n+1}b_{n+2}\phi_{n+2}
\label{eq:secondlattice}
\end{equation}
gives, for $b_n=\alpha n$,
\begin{equation}
\pd_t^2\phi=\alpha^2\Big[1+4(1+2\tilde x)\pd+4(1+\tilde x+\tilde x^2)\pd^2+\tfrac{8}{3}(1+2\tilde x)\pd^3+\tfrac43(1+\tilde x+\tilde x^2)\pd^4+\dots\Big]\phi .
\end{equation}
In terms of $x=\tilde x+\frac12$, $4(1+\tilde x+\tilde x^2)=4x^2+3$ and $\frac83(1+2\tilde x)=\frac{16}{3}x$, so the exact $O(x^{-2})$ part is $3\pd^2+\frac{16}{3}x\pd^3+\frac43x^2\pd^4$, of which the truncated result retains only $3\pd^2$.

\subsection{Log-depth form}
\label{app:logdepth}
Let $x=e^{U}$, $\Phi=x^{1/2}\phi$, $\cD=\pd_U$ and $\theta=x\pd_x$. Using $x^k\pd_x^k=\theta(\theta-1)\cdots(\theta-k+1)$ and $x^{1/2}\,x^{-m}P(\theta)\,x^{-1/2}=x^{-m}P(\cD-\frac12)$, the terms of eq.~\eqref{eq:Lexp} contribute, for $B=\alpha x+\beta_1x^{-1}+\beta_2x^{-2}+\dots$:
\begin{align}
-(2\alpha x\pd_x+\alpha)\ &\to\ -2\alpha\cD,\nonumber\\
-(2\beta_1x^{-1}\pd_x-\beta_1x^{-2})\ &\to\ -2\beta_1\,e^{-2U}(\cD-1),\nonumber\\
-\alpha\big(\tfrac13x\pd_x^3+\tfrac12\pd_x^2\big)\ &\to\ -\tfrac{\alpha}{3}\,e^{-2U}\big(\cD-\tfrac12\big)\big(\cD-\tfrac32\big)\big(\cD-1\big),
\end{align}
while $\beta_2$, the third-order terms with $\beta_1$, and the fifth-order terms with $\alpha x$ contribute at $O(e^{-3U})$ and $O(e^{-4U})$. Since $(\cD-\frac12)(\cD-\frac32)=\cD^2-2\cD+\frac34$, this gives eq.~\eqref{eq:subleading}. For the second-order equation, $\pd_t^2\Phi=\alpha^2[2\cD+e^{-2U}R][2\cD+e^{-2U}R]\Phi$, and $\cD e^{-2U}=e^{-2U}(\cD-2)$ gives $V_{\rm K}=2(\cD-2)R+2R\cD=4(\cD-1)R$. If the first correction is at $O(e^{-pU})$ the same computation gives $V_p=2(2\cD-p)R_p$. For the $\SL$ chain, $B=\alpha\sqrt{x^2-\nu^2}=\alpha x-\frac{\alpha\nu^2}{2x}+O(x^{-3})$ with $\nu=h-\frac12$, and
\begin{align}
\pd_t\Phi&=-\alpha\Big[2\cD+\tfrac13e^{-2U}(\cD-1)\big(\cD(\cD-2)-3h(h-1)\big)\Big]\Phi+O(e^{-4U}),\label{eq:first}\\
\pd_t^2\Phi&=\alpha^2\Big[4\cD^2+\tfrac43e^{-2U}(\cD-1)^2\big(\cD(\cD-2)-3h(h-1)\big)\Big]\Phi+O(e^{-4U}).\label{eq:second}
\end{align}
Numerically, the residual of the exact lattice action after subtracting eq.~\eqref{eq:first} scales as $x^{-4}$ for $h=\frac12,0.8,1.7$, and the lattice recursion of Proposition~\ref{prop:poly} reproduces $c_2=R(ik)/4$ for several values of $\beta_1$.

%
\section{Klein--Gordon equation near the AdS\texorpdfstring{$_2$}{2} horizon}
\label{app:KG}
\subsection{Exact form in the coordinate \texorpdfstring{$\zeta$}{zeta}}
For the metric~\eqref{eq:metric}, $\sqrt{-g}=2Le^{-2\zeta}$ and $\Box\Phi=e^{2\zeta}[-F^{-1}\pd_t^2\Phi+(2L)^{-2}\pd_\zeta(F\pd_\zeta\Phi)]$. Hence, exactly,
\begin{equation}
\pd_t^2\Phi=\frac{F^2}{4L^2}\pd_\zeta^2\Phi+\frac{F\,\pd_\zeta F}{4L^2}\pd_\zeta\Phi-m^2Fe^{-2\zeta}\Phi,
\end{equation}
i.e., using $F_0^2/4L^2=(2\pi T)^2$,
\begin{equation}
\pd_t^2\Phi=(2\pi T)^2\Big[\Big(1+\frac{2}{F_0}e^{-2\zeta}+\frac{e^{-4\zeta}}{F_0^2}\Big)\pd_\zeta^2-\frac{2}{F_0}\Big(e^{-2\zeta}+\frac{e^{-4\zeta}}{F_0}\Big)\pd_\zeta-\frac{4m^2L^2}{F_0}\Big(e^{-2\zeta}+\frac{e^{-4\zeta}}{F_0}\Big)\Big]\Phi ,
\label{eq:KGexact}
\end{equation}
whose expansion to $O(e^{-2\zeta})$ is eq.~\eqref{eq:KG}. The term $(2/F_0)e^{-2\zeta}\pd_\zeta^2$ is of the same order as the other corrections. With $\pd_\zeta=\cD$ and $(2\pi T)^2=4\alpha^2$ this gives $V_{\rm G}$ in eq.~\eqref{eq:VKVG}. The tortoise coordinate is
\begin{equation}
r_*=\frac{1}{4\pi T}\ln\frac{r-r_h}{r+r_h}=-\frac{1}{4\pi T}\ln\big(1+F_0e^{2\zeta}\big),\qquad \zeta=-2\pi Tr_*-\tfrac12\ln F_0+O(e^{-2\zeta}).
\end{equation}

\subsection{Exact ingoing solution and special points}
\label{app:exactKG}
With $\Phi=e^{-i\omega t}R(r)$, $X=r/r_h$ and $w=\omega/2\pi T=\omega L^2/r_h$, the Klein--Gordon equation becomes the associated Legendre equation
\begin{equation}
(1-X^2)R''-2XR'+\Big[m^2L^2+\frac{w^2}{1-X^2}\Big]R=0,
\end{equation}
with degree $\nu(\nu+1)=m^2L^2$ and order $\mu=\pm iw$. The ingoing solution, $R\propto e^{-i\omega r_*}$ near the horizon, is
\begin{equation}
R_{\rm in}=\Big(\frac{X+1}{X-1}\Big)^{iw/2}{}_2F_1\Big(1-\Delta,\Delta;1-iw;\frac{1-X}{2}\Big),
\end{equation}
with $\Delta(\Delta-1)=m^2L^2$. 
Continued to the boundary, $R_{\rm in}\simeq A\,(\frac{X-1}{2})^{\Delta-1}+B\,(\frac{X-1}{2})^{-\Delta}$ with $A=\Gamma(1-iw)\Gamma(2\Delta-1)/[\Gamma(\Delta)\Gamma(\Delta-iw)]$, and the normalizable condition $A=0$ gives the quasinormal frequencies $w=-i(\Delta+k)$; the factor $\Gamma(1-iw)$ cancels in $G_R\propto B/A$. Here we need instead the expansion of $R_{\rm in}$ near the horizon. Expanding in $y=X-1$, $R_{\rm in}=2^{iw/2}y^{-iw/2}[1+a_1y+O(y^2)]$ with
\begin{equation}
a_1=\frac{iw}{4}-\frac{(1-\Delta)\Delta}{2(1-iw)}=\frac{w^2+iw+2m^2L^2}{4(1-iw)},
\end{equation}
which is eq.~\eqref{eq:a1}. Since $y=(r-r_h)/r_h=(L/r_h)e^{-2\zeta}$ and $y^{-iw/2}\propto e^{iw\zeta}$, $a_1=c(k)\,r_h/L$ with $c(k)$ computed from $V_{\rm G}$ via eq.~\eqref{eq:ck}. The coefficients of ${}_2F_1$ in powers of $y$ are $(1-\Delta)_j(\Delta)_j/[(1-iw)_j\,j!\,(-2)^j]$, with poles at $iw=1,2,\dots,j$, i.e., at the special points $\omega_n=-2\pi iTn$. For $\Delta=1$ all of them vanish, and $R_{\rm in}=2^{iw/2}y^{-iw/2}(1+y/2)^{iw/2}$, whose expansion coefficients are polynomials in $w$.

%
\section{\texorpdfstring{$\SL$}{SL(2,R)} structure}
\label{app:sl2}
\paragraph{Bulk.} The vector fields~\eqref{eq:killing} satisfy the Killing equation for $-f\,dt^2+f^{-1}dr^2$ with $f=(r^2-r_h^2)/L^2$. Writing $\kappa=2\pi T=r_h/L^2$, $\xi_{\pm1}=e^{\mp\kappa t}(A\kappa^{-1}\pd_t\pm\mathcal{C}\pd_r)$ with $A=r/\sqrt{r^2-r_h^2}$ and $\mathcal{C}=\sqrt{r^2-r_h^2}$, one finds $[\xi_1,\xi_{-1}]=2(A^2+\mathcal{C}A')\xi_0=2\xi_0$, and
\begin{equation}
C=(1-A^2)\kappa^{-2}\pd_t^2+\mathcal{C}^2\pd_r^2+(\mathcal{C}\mathcal{C}'+A\mathcal{C})\pd_r=-\frac{L^2}{f}\pd_t^2+L^2\pd_r(f\pd_r)=L^2\Box .
\end{equation}

\paragraph{Krylov side.} On $\Dser{h}$, $[L_0,L_{\pm1}]=\mp L_{\pm1}$ and $[L_1,L_{-1}]=2L_0$. The anti-Hermitian elements $iL_0$ (compact) and $i\cB$, $\frac12(L_1-L_{-1})$ (non-compact) span the real Lie algebra. One finds $[\cB,L_0]=\frac12(L_1-L_{-1})$ and $[\cB,\frac12(L_1-L_{-1})]=-L_0$, so ${\rm ad}_{i\cB}$ maps $iL_0\mapsto-\frac12(L_1-L_{-1})$ and $\frac12(L_1-L_{-1})\mapsto-iL_0$, and its eigenvalues are $\pm1$ and $0$, as for $\xi_0$. The coherent state $e^{2i\alpha\cB t}|h,0\rangle=\sum_n i^n\phi_n(t)|h,n\rangle$ reproduces eq.~\eqref{eq:sl2sol}, with $\phi_0=\cosh^{-2h}(\alpha t)$ and $\langle L_0\rangle=h\cosh2\alpha t$.

\paragraph{CFT$_1$ check.}
\label{app:cft1}
The spectral density of $\cosh^{-2\Delta}(\pi Tt)$ is $\propto|\Gamma(\Delta+i\omega/2\pi T)|^2$. The Stieltjes procedure applied to this density reproduces $b_n=\pi T\sqrt{n(n+2\Delta-1)}$ to machine precision ($n\le25$, $\Delta=0.25,0.5,1,2.3$). Its poles in the lower half-plane, $\omega=-2\pi iT(\Delta+k)$, $k=0,1,\dots$, are the quasinormal frequencies of the AdS$_2$ black hole~\cite{Faulkner:2009wj,Birmingham:2001pj}.

\paragraph{Generalized eigenfunctions.} The generalized eigenfunctions of the $\SL$ chain are Meixner-Pollaczek functions~\cite{Muck:2022xfc}, with generating function $\sum_nP_n\,u^n=(1-iu)^{-h+iX}(1+iu)^{-h-iX}$, $X=\Omega/2\alpha$, where $\Omega$ is the eigenvalue of $\mathcal{L}$, so that $\phi\propto e^{i\Omega t}$ and $k=-X$. Darboux's method gives the asymptotic expansion of the smooth component as $\sum_j d_j(X)\,\Gamma(n+h-iX-j)/[\Gamma(h-iX-j)\,n!]$ with $d_j$ polynomial in $X$ up to an overall factor; every coefficient of its large-$n$ expansion is therefore entire in the frequency, at all orders, an independent check of Proposition~\ref{prop:poly} for this chain. In the variable $x=n+h$ the $1/x$ term cancels identically, and the $x^{-2}$ coefficient reproduces $c(k)$ of eq.~\eqref{eq:ck} with $V_{\rm K}$, $c_{\rm K}(k)=\frac{1}{12}(ik-1)\big(ik(ik-2)-3h(h-1)\big)$.

%
\section{Formal solutions and invariants}
\label{app:formal}
\subsection{Action of the allowed transformations}
\label{app:transf}
(a) For $\Phi=(1+zQ)\tilde\Phi$, $z=e^{-2U}$, multiplying the equation by $(1-zQ)$ gives $\pd_t^2\tilde\Phi=\alpha^2[4\cD^2+zV+4(\cD^2zQ-zQ\cD^2)]\tilde\Phi+O(z^2)$, and $\cD^2z=z(\cD-2)^2$ gives $V\to V-16(\cD-1)Q$. (b) For $U=\tilde U+\lambda_1e^{-2\tilde U}$, $\pd_U=(1+2\lambda_1z)\pd_{\tilde U}+O(z^2)$, and $4\cD^2\to4\tilde\cD^2+16\lambda_1z\tilde\cD(\tilde\cD-1)$. (c) For $t=\tilde t+G(U)$, a solution $e^{-i\omega t}\Phi_\omega(U)$ becomes $e^{-i\omega\tilde t}e^{-i\omega G(U)}\Phi_\omega(U)$; since $G$ is analytic in $z$, the factor $e^{-i\omega G}$ has an expansion in $z$ with coefficients entire in $\omega$. (d) Multiplication of the equation by an invertible operator does not change its solutions. A constant shift $U\to U+u_0$ multiplies $c(k)$ by $e^{-2u_0}$ and the solution by the entire phase $e^{iku_0}$. In all cases, the coefficients of the normalized ingoing solution change by entire functions of $\omega$ (polynomials in (a), (b), (d)), so the presence of a pole at a special point, and at the first order its residue up to a positive factor, are invariant.

\subsection{Second-order recursion}
For the second-order lattice equation~\eqref{eq:secondlattice}, substituting $\phi=\sum_qc_qx^{s-q}$ and collecting the power $x^{s-Q}$ gives a recursion whose coefficient of $c_Q$ is $4\alpha^2[(ik-Q)^2-(ik)^2]=4\alpha^2Q(Q-2ik)$, from $\Lk^2\simeq4\alpha^2\cD^2$ acting on $x^{s-Q}$ and the term $\omega^2$. The powers $x^{s+2}$ and $x^{s+1}$ vanish identically.

\subsection{Massless bulk field to all orders}
\label{app:massless}
For $m=0$, $R_{\rm in}\propto y^{-iw/2}(1+y/2)^{iw/2}$, whose coefficients are polynomials in $w$ (appendix~\ref{app:exactKG}). Let $\Phi^{(1)}=e^{ikU}\sum_Qc^{(1)}_Qz^{Q/2}$ and $\Phi^{(2)}=e^{ikU}\sum_Qc^{(2)}_Qz^{Q/2}$ be two formal ingoing solutions with $c^{(1)}_0=c^{(2)}_0=1$ and coefficients polynomial in $k$. Since $z^{R/2}e^{ikU}=e^{(ik-R)U}$ is an eigenfunction of $\cD$ with eigenvalue $ik-R$, a field redefinition $\Phi^{(1)}=[1+\sum_{P\ge1}z^{P/2}Q_P(\cD)]\Phi^{(2)}$ maps one onto the other provided, order by order,
\begin{equation}
Q_Q(ik)=c^{(1)}_Q-c^{(2)}_Q-\sum_{P=1}^{Q-1}Q_P\big(ik-Q+P\big)\,c^{(2)}_{Q-P}.
\end{equation}
By induction every $Q_Q$ is a polynomial, so the redefinition is local and independent of $\omega$. This applies to the smooth Krylov chain and the massless bulk field, but not to a massive one, for which $c^{(2)}_2$ has a pole.

\subsection{Staggered Lanczos component}
\label{app:staggered}
Write $b_n=\bar b_n+(-1)^ne_n$ and $\phi_n=A_n+(-1)^nS_n$ with $\bar b$, $e$, $A$, $S$ smooth. Separating smooth and staggered parts of eq.~\eqref{eq:chain} with $\pd_t\to-i\omega$,
\begin{align}
-i\omega A_n&=\bar b_nA_{n-1}-\bar b_{n+1}A_{n+1}-e_nS_{n-1}-e_{n+1}S_{n+1},\nonumber\\
-i\omega S_n&=e_nA_{n-1}+e_{n+1}A_{n+1}-\bar b_nS_{n-1}+\bar b_{n+1}S_{n+1}.
\end{align}
The staggered sector obeys the time-reversed chain, with free solutions $S\propto x^{-ik-\frac12}$ (outgoing). For $\bar b_n=\alpha(x-\frac12)$ (link-centered) and $e_n=\varepsilon\alpha(x-\frac12)^{-\gamma}$, the ingoing $A=x^{s}$, $s=ik-\frac12$, sources $S=\sigma x^{s-\gamma}$ with $[-i\omega-2\alpha(s-\gamma)-\alpha]\sigma=2\varepsilon\alpha$ at leading order, i.e., $\sigma=\varepsilon/(\gamma-2ik)$, eq.~\eqref{eq:sigma}. 

%
\section{BTZ details}
\label{app:btz}
\subsection{Spectral density at finite momentum}
\label{app:btzft}
In light-cone coordinates $u=x-t$, $v=x+t$, the Wightman correlator of section~\ref{sec:btz} is the product $[\pi T/\cosh\pi Tu]^{\Delta}\,[\pi T/\cosh\pi Tv]^{\Delta}$, and
\begin{equation}
\omega t-qx=-\frac{\omega+q}{2}\,u+\frac{\omega-q}{2}\,v,\qquad dt\,dx=\tfrac12\,du\,dv .
\end{equation} The Fourier transform therefore factorizes. Each factor is the transform of the $\SL$ autocorrelation of section~\ref{sec:sl2chain} with $h=\Delta/2$ and $\alpha=\pi T$, evaluated at frequency $k=(\omega\pm q)/2$: $\int dy\,e^{iky}\cosh^{-\Delta}(\pi Ty)\propto|\Gamma(\frac\Delta2+\frac{ik}{2\pi T})|^2$. This gives eq.~\eqref{eq:rhobtz}.

\subsection{Continuous Hahn recurrence}
\label{app:btzhahn}
The continuous Hahn polynomials obey~\cite{Koekoek_2010}
\begin{equation}
(a+ix)\tilde p_n=A_n\tilde p_{n+1}-(A_n+C_n)\tilde p_n+C_n\tilde p_{n-1},
\end{equation}
\begin{equation}
A_n=-\frac{(n+\Sigma-1)(n+a+c)(n+a+d)}{(2n+\Sigma-1)(2n+\Sigma)},\qquad
C_n=\frac{n(n+b+c-1)(n+b+d-1)}{(2n+\Sigma-2)(2n+\Sigma-1)},
\end{equation}
with $\Sigma=a+b+c+d$. Rescaling to monic polynomials in $y=a+ix$ and then in $x$ gives $xR_n=R_{n+1}+i(a+A_n+C_n)R_n-A_{n-1}C_nR_{n-1}$, so the Lanczos coefficients in the variable $x$ are $b_n^2=-A_{n-1}C_n$, and the diagonal coefficients $i(a+A_n+C_n)$ vanish for the symmetric weight. With $\Sigma=2\Delta$, $a+c=b+d=\Delta$, $a+d=\Delta+2is$ and $b+c=\Delta-2is$, $s=q/4\pi T$,
\begin{equation}\label{app:e4}
b_n^2=\frac{n(n+2\Delta-2)(n+\Delta-1)^2\big[(n+\Delta-1)^2+4s^2\big]}{(2n+2\Delta-3)(2n+2\Delta-2)^2(2n+2\Delta-1)}=\frac{\big(m^2-(\Delta-1)^2\big)\big(m^2+4s^2\big)}{4(4m^2-1)},
\end{equation}
with $m=n+\Delta-1$.
Since $\omega=4\pi Tx$, the monic polynomials in $\omega$ are $(4\pi T)^nR_n(x)$, so the Lanczos coefficients in $\omega$ are $4\pi T$ times those in $x$; with $4s^2=(q/2\pi T)^2=Q^2$, eq.~\eqref{app:e4} becomes eq.~\eqref{eq:bnbtz}.

\subsection{Bulk Frobenius expansion and pole-skipping}
\label{app:btzbulk}
The BTZ metric $ds^2=-(r^2-r_h^2)dt^2+dr^2/(r^2-r_h^2)+r^2dx^2$ has $T=r_h/2\pi$ and $\sqrt{-g}=r$, which gives eq.~\eqref{eq:btzradial}. Multiplying it by $\varepsilon$ and expanding, a term $a_j\varepsilon^{\sigma}$ contributes $(2\sigma^2+\frac{w^2}2)a_j$ at order $\varepsilon^{\sigma}$ and $(\sigma^2+3\sigma-\frac{w^2}4-Q^2-m^2)a_j$ at order $\varepsilon^{\sigma+1}$. The indicial equation is $2\sigma^2+w^2/2=0$; the ingoing exponent is $\sigma_0=-iw/2$, and $a_1=-(\sigma_0^2+3\sigma_0-\frac{w^2}4-Q^2-m^2)/(4\sigma_0+2)$, which is eq.~\eqref{eq:btzres}. The exact retarded function~\cite{Son:2002sd}
\begin{equation}
G_R\propto\frac{\Gamma\big(\frac\Delta2-\frac{i(w-Q)}2\big)\Gamma\big(\frac\Delta2-\frac{i(w+Q)}2\big)}{\Gamma\big(1-\frac\Delta2-\frac{i(w-Q)}2\big)\Gamma\big(1-\frac\Delta2-\frac{i(w+Q)}2\big)}
\end{equation}
is ill-defined at $w=-i$, $Q=\pm i(\Delta-1)$, where a pole of the numerator meets a pole of the denominator (pole-skipping); these are the momenta at which the residue~\eqref{eq:btzres} vanishes.

%

\begin{thebibliography}{10}

\bibitem{liesen2012krylov}
J.~Liesen and Z.~Strakos, \emph{Krylov subspace methods principles and
  analysis}.
\newblock Numerical Mathematics and Scientific Computation. Oxford University
  Press, Oxford, 2012.

\bibitem{Viswanath_1994}
V.~S. Viswanath and G.~M{\"u}ller, \emph{The Recursion Method: Application to
  Many-Body Dynamics}.
\newblock Springer Berlin Heidelberg, 1994,
  \href{http://dx.doi.org/10.1007/978-3-540-48651-0}{10.1007/978-3-540-48651-0}.

\bibitem{krylov1931}
A.~Krylov, \emph{De la r\'esolution num\'erique de l'\'equation servant \`a
  d\'eterminer dans des questions de m\'ecanique appliqu\'ee les fr\'equences
  de petites oscillations des syst\`emes mat\'eriels},
  \href{http://dx.doi.org/10.1017/CBO9781139424400}{\emph{Bulletin de
  l'Acad\'emie des Sciences de l'URSS. Classe des sciences math\'ematiques et
  naturelles} {\bf 4} (1931) 491--539}.

\bibitem{lanczos1950}
C.~Lanczos, \emph{An iteration method for the solution of the eigenvalue
  problem of linear differential and integral operators}, {\emph{Journal of
  research of the National Bureau of Standards} {\bf 45} (1950) 255--282}.

\bibitem{Mattis1981}
D.~C. Mattis, \emph{{How to Reduce Practically Any Problem to One Dimension}},
  in \emph{Physics in One Dimension}, pp.~3--10.
\newblock Springer, Berlin, Heidelberg, 1981.
\newblock \href{http://dx.doi.org/10.1007/978-3-642-81592-8\_1}{DOI}.

\bibitem{Parker:2018yvk}
D.~E. Parker, X.~Cao, A.~Avdoshkin, T.~Scaffidi and E.~Altman, \emph{{A
  Universal Operator Growth Hypothesis}},
  \href{http://dx.doi.org/10.1103/PhysRevX.9.041017}{\emph{Phys. Rev. X} {\bf
  9} (2019) 041017}, [\href{http://arxiv.org/abs/1812.08657}{{\tt
  1812.08657}}].

\bibitem{Dymarsky:2019elm}
A.~Dymarsky and A.~Gorsky, \emph{{Quantum chaos as delocalization in Krylov
  space}}, \href{http://dx.doi.org/10.1103/PhysRevB.102.085137}{\emph{Phys.
  Rev. B} {\bf 102} (2020) 085137},
  [\href{http://arxiv.org/abs/1912.12227}{{\tt 1912.12227}}].

\bibitem{Balasubramanian:2022tpr}
V.~Balasubramanian, P.~Caputa, J.~M. Magan and Q.~Wu, \emph{{Quantum chaos and
  the complexity of spread of states}},
  \href{http://dx.doi.org/10.1103/PhysRevD.106.046007}{\emph{Phys. Rev. D} {\bf
  106} (2022) 046007}, [\href{http://arxiv.org/abs/2202.06957}{{\tt
  2202.06957}}].

\bibitem{Caputa:2024vrn}
P.~Caputa, H.-S. Jeong, S.~Liu, J.~F. Pedraza and L.-C. Qu, \emph{{Krylov
  complexity of density matrix operators}},
  \href{http://dx.doi.org/10.1007/JHEP05(2024)337}{\emph{JHEP} {\bf 05} (2024)
  337}, [\href{http://arxiv.org/abs/2402.09522}{{\tt 2402.09522}}].

\bibitem{Nandy:2024evd}
P.~Nandy, A.~S. Matsoukas-Roubeas, P.~Mart{\'\i}nez-Azcona, A.~Dymarsky and
  A.~del Campo, \emph{{Quantum dynamics in Krylov space: Methods and
  applications}},
  \href{http://dx.doi.org/10.1016/j.physrep.2025.05.001}{\emph{Phys. Rept.}
  {\bf 1125-1128} (2025) 1--82}, [\href{http://arxiv.org/abs/2405.09628}{{\tt
  2405.09628}}].

\bibitem{Rabinovici:2025otw}
E.~Rabinovici, A.~S{\'a}nchez-Garrido, R.~Shir and J.~Sonner, \emph{{Krylov
  Complexity}},  \href{http://arxiv.org/abs/2507.06286}{{\tt 2507.06286}}.

\bibitem{Jeong:2026gdc}
H.-S. Jeong and J.~F. Pedraza, \emph{{Quantum Chaos and Spread of States in
  Krylov Subspace: A Topical Review}},
  \href{http://arxiv.org/abs/2608.26825}{{\tt 2608.26825}}.

\bibitem{Bharti:2021aa}
K.~Bharti and T.~Haug, \emph{{Iterative quantum-assisted eigensolver}},
  \href{http://dx.doi.org/10.1103/PhysRevA.104.L050401}{\emph{Phys. Rev. A}
  {\bf 104} (2021) L050401}, [\href{http://arxiv.org/abs/2010.05638}{{\tt
  2010.05638}}].

\bibitem{Cortes:2022aa}
C.~L. Cortes and S.~K. Gray, \emph{{Quantum Krylov subspace algorithms for
  ground- and excited-state energy estimation}}, {\emph{Phys. Rev. A} {\bf 105}
  (2022) }, [\href{http://arxiv.org/abs/2109.06868}{{\tt 2109.06868}}].

\bibitem{Kirby2023exactefficient}
W.~Kirby, M.~Motta and A.~Mezzacapo, \emph{Exact and efficient {L}anczos method
  on a quantum computer},
  \href{http://dx.doi.org/10.22331/q-2023-05-23-1018}{\emph{{Quantum}} {\bf 7}
  (May, 2023) 1018}.

\bibitem{Bhattacharya2022}
A.~Bhattacharya, P.~Nandy, P.~P. Nath and H.~Sahu, \emph{{Operator growth and
  Krylov construction in dissipative open quantum systems}}, {\emph{JHEP} {\bf
  12} (2022) 081}, [\href{http://arxiv.org/abs/2207.05347}{{\tt 2207.05347}}].

\bibitem{Liu:2022god}
C.~Liu, H.~Tang and H.~Zhai, \emph{{Krylov complexity in open quantum
  systems}},
  \href{http://dx.doi.org/10.1103/PhysRevResearch.5.033085}{\emph{Phys. Rev.
  Res.} {\bf 5} (2023) 033085}, [\href{http://arxiv.org/abs/2207.13603}{{\tt
  2207.13603}}].

\bibitem{Bhattacharya2023}
A.~Bhattacharya, P.~Nandy, P.~P. Nath and H.~Sahu, \emph{{On Krylov complexity
  in open systems: an approach via bi-Lanczos algorithm}}, {\emph{JHEP} {\bf
  12} (2023) 066}, [\href{http://arxiv.org/abs/2303.04175}{{\tt 2303.04175}}].

\bibitem{Baggioli:2025knt}
M.~Baggioli, K.-B. Huh, H.-S. Jeong, X.~Jiang, K.-Y. Kim and J.~F. Pedraza,
  \emph{{Quantum Chaos Diagnostics for non-Hermitian Systems from Bi-Lanczos
  Krylov Dynamics}},  \href{http://arxiv.org/abs/2508.13956}{{\tt 2508.13956}}.

\bibitem{Avdoshkin:2019trj}
A.~Avdoshkin and A.~Dymarsky, \emph{{Euclidean operator growth and quantum
  chaos}},
  \href{http://dx.doi.org/10.1103/PhysRevResearch.2.043234}{\emph{Phys. Rev.
  Res.} {\bf 2} (2020) 043234}, [\href{http://arxiv.org/abs/1911.09672}{{\tt
  1911.09672}}].

\bibitem{Maldacena:2015waa}
J.~Maldacena, S.~H. Shenker and D.~Stanford, \emph{{A bound on chaos}},
  \href{http://dx.doi.org/10.1007/JHEP08(2016)106}{\emph{JHEP} {\bf 08} (2016)
  106}, [\href{http://arxiv.org/abs/1503.01409}{{\tt 1503.01409}}].

\bibitem{Dymarsky:2021bjq}
A.~Dymarsky and M.~Smolkin, \emph{{Krylov complexity in conformal field
  theory}}, \href{http://dx.doi.org/10.1103/PhysRevD.104.L081702}{\emph{Phys.
  Rev. D} {\bf 104} (2021) L081702},
  [\href{http://arxiv.org/abs/2104.09514}{{\tt 2104.09514}}].

\bibitem{Maldacena:1997re}
J.~M. Maldacena, \emph{{The Large N limit of superconformal field theories and
  supergravity}}, \href{http://dx.doi.org/10.1023/A:1026654312961,
  10.1023/A:1026654312961}{\emph{Adv.Theor.Math.Phys.} {\bf 2} (1998)
  231--252}, [\href{http://arxiv.org/abs/hep-th/9711200}{{\tt
  hep-th/9711200}}].

\bibitem{Gubser:1998bc}
S.~S. Gubser, I.~R. Klebanov and A.~M. Polyakov, \emph{{Gauge theory
  correlators from non-critical string theory}},
  \href{http://dx.doi.org/10.1016/S0370-2693(98)00377-3}{\emph{Phys. Lett.}
  {\bf B428} (1998) 105--114}, [\href{http://arxiv.org/abs/hep-th/9802109}{{\tt
  hep-th/9802109}}].

\bibitem{Witten:1998qj}
E.~Witten, \emph{{Anti-de Sitter space and holography}}, {\emph{Adv. Theor.
  Math. Phys.} {\bf 2} (1998) 253--291},
  [\href{http://arxiv.org/abs/hep-th/9802150}{{\tt hep-th/9802150}}].

\bibitem{Jian:2020qpp}
S.-K. Jian, B.~Swingle and Z.-Y. Xian, \emph{{Complexity growth of operators in
  the SYK model and in JT gravity}},
  \href{http://dx.doi.org/10.1007/JHEP03(2021)014}{\emph{JHEP} {\bf 03} (2021)
  014}, [\href{http://arxiv.org/abs/2008.12274}{{\tt 2008.12274}}].

\bibitem{Lin:2022rbf}
H.~W. Lin, \emph{{The bulk Hilbert space of double scaled SYK}},
  \href{http://dx.doi.org/10.1007/JHEP11(2022)060}{\emph{JHEP} {\bf 11} (2022)
  060}, [\href{http://arxiv.org/abs/2208.07032}{{\tt 2208.07032}}].

\bibitem{Rabinovici:2023yex}
E.~Rabinovici, A.~S\'anchez-Garrido, R.~Shir and J.~Sonner, \emph{{A bulk
  manifestation of Krylov complexity}},
  \href{http://dx.doi.org/10.1007/JHEP08(2023)213}{\emph{JHEP} {\bf 08} (2023)
  213}, [\href{http://arxiv.org/abs/2305.04355}{{\tt 2305.04355}}].

\bibitem{Heller:2024ldz}
M.~P. Heller, J.~Papalini and T.~Schuhmann, \emph{{Krylov Spread Complexity as
  Holographic Complexity beyond Jackiw-Teitelboim Gravity}},
  \href{http://dx.doi.org/10.1103/spcr-jgm6}{\emph{Phys. Rev. Lett.} {\bf 135}
  (2025) 151602}, [\href{http://arxiv.org/abs/2412.17785}{{\tt 2412.17785}}].

\bibitem{Ambrosini:2024sre}
M.~Ambrosini, E.~Rabinovici, A.~S{\'a}nchez-Garrido, R.~Shir and J.~Sonner,
  \emph{{Operator K-complexity in DSSYK: Krylov complexity equals bulk
  length}}, \href{http://dx.doi.org/10.1007/JHEP08(2025)059}{\emph{JHEP} {\bf
  08} (2025) 059}, [\href{http://arxiv.org/abs/2412.15318}{{\tt 2412.15318}}].

\bibitem{Fu:2025kkh}
Y.~Fu, H.-S. Jeong, K.-Y. Kim and J.~F. Pedraza, \emph{{Toward Krylov-based
  holography in double-scaled SYK}},
  \href{http://dx.doi.org/10.1007/JHEP05(2026)056}{\emph{JHEP} {\bf 05} (2026)
  056}, [\href{http://arxiv.org/abs/2510.22658}{{\tt 2510.22658}}].

\bibitem{Susskind:2014rva}
L.~Susskind, \emph{{Computational Complexity and Black Hole Horizons}},
  \href{http://dx.doi.org/10.1002/prop.201500093,
  10.1002/prop.201500092}{\emph{Fortsch. Phys.} {\bf 64} (2016) 44--48},
  [\href{http://arxiv.org/abs/1403.5695}{{\tt 1403.5695}}].

\bibitem{Stanford:2014jda}
D.~Stanford and L.~Susskind, \emph{{Complexity and Shock Wave Geometries}},
  \href{http://dx.doi.org/10.1103/PhysRevD.90.126007}{\emph{Phys. Rev.} {\bf
  D90} (2014) 126007}, [\href{http://arxiv.org/abs/1406.2678}{{\tt
  1406.2678}}].

\bibitem{Chapman:2021jbh}
S.~Chapman and G.~Policastro, \emph{{Quantum computational complexity from
  quantum information to black holes and back}},
  \href{http://dx.doi.org/10.1140/epjc/s10052-022-10037-1}{\emph{Eur. Phys. J.
  C} {\bf 82} (2022) 128}, [\href{http://arxiv.org/abs/2110.14672}{{\tt
  2110.14672}}].

\bibitem{Chen:2021lnq}
B.~Chen, B.~Czech and Z.-z. Wang, \emph{{Quantum information in holographic
  duality}}, \href{http://dx.doi.org/10.1088/1361-6633/ac51b5}{\emph{Rept.
  Prog. Phys.} {\bf 85} (2022) 046001},
  [\href{http://arxiv.org/abs/2108.09188}{{\tt 2108.09188}}].

\bibitem{Sekino:2008he}
Y.~Sekino and L.~Susskind, \emph{{Fast Scramblers}},
  \href{http://dx.doi.org/10.1088/1126-6708/2008/10/065}{\emph{JHEP} {\bf 10}
  (2008) 065}, [\href{http://arxiv.org/abs/0808.2096}{{\tt 0808.2096}}].

\bibitem{Susskind:2018tei}
L.~Susskind, \emph{{Why do Things Fall?}},
  \href{http://arxiv.org/abs/1802.01198}{{\tt 1802.01198}}.

\bibitem{Brown:2018kvn}
A.~R. Brown, H.~Gharibyan, A.~Streicher, L.~Susskind, L.~Thorlacius and
  Y.~Zhao, \emph{{Falling Toward Charged Black Holes}},
  \href{http://dx.doi.org/10.1103/PhysRevD.98.126016}{\emph{Phys. Rev. D} {\bf
  98} (2018) 126016}, [\href{http://arxiv.org/abs/1804.04156}{{\tt
  1804.04156}}].

\bibitem{Lin:2019qwu}
H.~W. Lin, J.~Maldacena and Y.~Zhao, \emph{{Symmetries Near the Horizon}},
  \href{http://dx.doi.org/10.1007/JHEP08(2019)049}{\emph{JHEP} {\bf 08} (2019)
  049}, [\href{http://arxiv.org/abs/1904.12820}{{\tt 1904.12820}}].

\bibitem{Magan:2018nmu}
J.~M. Mag{\'a}n, \emph{{Black holes, complexity and quantum chaos}},
  \href{http://dx.doi.org/10.1007/JHEP09(2018)043}{\emph{JHEP} {\bf 09} (2018)
  043}, [\href{http://arxiv.org/abs/1805.05839}{{\tt 1805.05839}}].

\bibitem{Kar:2021nbm}
A.~Kar, L.~Lamprou, M.~Rozali and J.~Sully, \emph{{Random matrix theory for
  complexity growth and black hole interiors}},
  \href{http://dx.doi.org/10.1007/JHEP01(2022)016}{\emph{JHEP} {\bf 01} (2022)
  016}, [\href{http://arxiv.org/abs/2106.02046}{{\tt 2106.02046}}].

\bibitem{Iliesiu:2021ari}
L.~V. Iliesiu, M.~Mezei and G.~S{\'a}rosi, \emph{{The volume of the black hole
  interior at late times}},
  \href{http://dx.doi.org/10.1007/JHEP07(2022)073}{\emph{JHEP} {\bf 07} (2022)
  073}, [\href{http://arxiv.org/abs/2107.06286}{{\tt 2107.06286}}].

\bibitem{Dodelson:2025rng}
M.~Dodelson, \emph{{Black holes from chaos}},
  \href{http://arxiv.org/abs/2501.06170}{{\tt 2501.06170}}.

\bibitem{Kubo_1957}
R.~Kubo, \emph{Statistical-mechanical theory of irreversible processes. i.
  general theory and simple applications to magnetic and conduction problems},
  \href{http://dx.doi.org/10.1143/jpsj.12.570}{\emph{Journal of the Physical
  Society of Japan} {\bf 12} (1957) 570--586}.

\bibitem{Martin:1959aa}
P.~C. Martin and J.~Schwinger, \emph{{Theory of many-particle systems. I}},
  \href{http://dx.doi.org/10.1103/PhysRev.115.1342}{\emph{Phys. Rev.} {\bf 115}
  (1959) 1342--1373}.

\bibitem{Breitenlohner:1982bm}
P.~Breitenlohner and D.~Z. Freedman, \emph{{Positive Energy in anti-De Sitter
  Backgrounds and Gauged Extended Supergravity}},
  \href{http://dx.doi.org/10.1016/0370-2693(82)90643-8}{\emph{Phys. Lett. B}
  {\bf 115} (1982) 197--201}.

\bibitem{Breitenlohner:1982jf}
P.~Breitenlohner and D.~Z. Freedman, \emph{{Stability in Gauged Extended
  Supergravity}},
  \href{http://dx.doi.org/10.1016/0003-4916(82)90116-6}{\emph{Annals Phys.}
  {\bf 144} (1982) 249}.

\bibitem{Blake:2019otz}
M.~Blake, R.~A. Davison and D.~Vegh, \emph{{Horizon constraints on holographic
  Green{\textquoteright}s functions}},
  \href{http://dx.doi.org/10.1007/JHEP01(2020)077}{\emph{JHEP} {\bf 01} (2020)
  077}, [\href{http://arxiv.org/abs/1904.12883}{{\tt 1904.12883}}].

\bibitem{Blake:2017ris}
M.~Blake, H.~Lee and H.~Liu, \emph{{A quantum hydrodynamical description for
  scrambling and many-body chaos}},
  \href{http://dx.doi.org/10.1007/JHEP10(2018)127}{\emph{JHEP} {\bf 10} (2018)
  127}, [\href{http://arxiv.org/abs/1801.00010}{{\tt 1801.00010}}].

\bibitem{Grozdanov:2017ajz}
S.~Grozdanov, K.~Schalm and V.~Scopelliti, \emph{{Black hole scrambling from
  hydrodynamics}},
  \href{http://dx.doi.org/10.1103/PhysRevLett.120.231601}{\emph{Phys. Rev.
  Lett.} {\bf 120} (2018) 231601}, [\href{http://arxiv.org/abs/1710.00921}{{\tt
  1710.00921}}].

\bibitem{Blake:2018leo}
M.~Blake, R.~A. Davison, S.~Grozdanov and H.~Liu, \emph{{Many-body chaos and
  energy dynamics in holography}},
  \href{http://dx.doi.org/10.1007/JHEP10(2018)035}{\emph{JHEP} {\bf 10} (2018)
  035}, [\href{http://arxiv.org/abs/1809.01169}{{\tt 1809.01169}}].

\bibitem{Ahn:2025qat}
Y.~Ahn, S.~Grozdanov, H.-S. Jeong and J.~F. Pedraza, \emph{{Cosmological
  pole-skipping, shock waves and quantum chaotic dynamics of de Sitter
  horizons}}, \href{http://dx.doi.org/10.1007/JHEP10(2026)185}{\emph{JHEP} {\bf
  10} (2026) 185}, [\href{http://arxiv.org/abs/2508.15589}{{\tt 2508.15589}}].

\bibitem{Caputa:2021sib}
P.~Caputa, J.~M. Magan and D.~Patramanis, \emph{{Geometry of Krylov
  complexity}},
  \href{http://dx.doi.org/10.1103/PhysRevResearch.4.013041}{\emph{Phys. Rev.
  Res.} {\bf 4} (2022) 013041}, [\href{http://arxiv.org/abs/2109.03824}{{\tt
  2109.03824}}].

\bibitem{Maldacena:2016hyu}
J.~Maldacena and D.~Stanford, \emph{{Remarks on the Sachdev-Ye-Kitaev model}},
  \href{http://dx.doi.org/10.1103/PhysRevD.94.106002}{\emph{Phys. Rev. D} {\bf
  94} (2016) 106002}, [\href{http://arxiv.org/abs/1604.07818}{{\tt
  1604.07818}}].

\bibitem{Spradlin:1999bn}
M.~Spradlin and A.~Strominger, \emph{{Vacuum states for AdS(2) black holes}},
  \href{http://dx.doi.org/10.1088/1126-6708/1999/11/021}{\emph{JHEP} {\bf 11}
  (1999) 021}, [\href{http://arxiv.org/abs/hep-th/9904143}{{\tt
  hep-th/9904143}}].

\bibitem{Barbon:2019wsy}
J.~L.~F. Barb\'on, E.~Rabinovici, R.~Shir and R.~Sinha, \emph{{On The Evolution
  Of Operator Complexity Beyond Scrambling}},
  \href{http://dx.doi.org/10.1007/JHEP10(2019)264}{\emph{JHEP} {\bf 10} (2019)
  264}, [\href{http://arxiv.org/abs/1907.05393}{{\tt 1907.05393}}].

\bibitem{Jackiw:1984je}
R.~Jackiw, \emph{{Lower Dimensional Gravity}},
  \href{http://dx.doi.org/10.1016/0550-3213(85)90448-1}{\emph{Nucl. Phys. B}
  {\bf 252} (1985) 343--356}.

\bibitem{Teitelboim:1983ux}
C.~Teitelboim, \emph{{Gravitation and Hamiltonian Structure in Two Space-Time
  Dimensions}},
  \href{http://dx.doi.org/10.1016/0370-2693(83)90012-6}{\emph{Phys. Lett. B}
  {\bf 126} (1983) 41--45}.

\bibitem{Almheiri:2014cka}
A.~Almheiri and J.~Polchinski, \emph{{Models of AdS$_{2}$ backreaction and
  holography}}, \href{http://dx.doi.org/10.1007/JHEP11(2015)014}{\emph{JHEP}
  {\bf 11} (2015) 014}, [\href{http://arxiv.org/abs/1402.6334}{{\tt
  1402.6334}}].

\bibitem{Maldacena:2016upp}
J.~Maldacena, D.~Stanford and Z.~Yang, \emph{{Conformal symmetry and its
  breaking in two dimensional Nearly Anti-de-Sitter space}},
  \href{http://dx.doi.org/10.1093/ptep/ptw124}{\emph{PTEP} {\bf 2016} (2016)
  12C104}, [\href{http://arxiv.org/abs/1606.01857}{{\tt 1606.01857}}].

\bibitem{Strominger:1998yg}
A.~Strominger, \emph{{AdS(2) quantum gravity and string theory}},
  \href{http://dx.doi.org/10.1088/1126-6708/1999/01/007}{\emph{JHEP} {\bf 01}
  (1999) 007}, [\href{http://arxiv.org/abs/hep-th/9809027}{{\tt
  hep-th/9809027}}].

\bibitem{Mertens:2022irh}
T.~G. Mertens and G.~J. Turiaci, \emph{{Solvable models of quantum black holes:
  a review on Jackiw{\textendash}Teitelboim gravity}},
  \href{http://dx.doi.org/10.1007/s41114-023-00046-1}{\emph{Living Rev. Rel.}
  {\bf 26} (2023) 4}, [\href{http://arxiv.org/abs/2210.10846}{{\tt
  2210.10846}}].

\bibitem{Kitaev:2017hnr}
A.~Kitaev, \emph{{Notes on $\widetilde{\mathrm{SL}}(2,\mathbb{R})$
  representations}},  \href{http://arxiv.org/abs/1711.08169}{{\tt 1711.08169}}.

\bibitem{Caputa:2021ori}
P.~Caputa and S.~Datta, \emph{{Operator growth in 2d CFT}},
  \href{http://dx.doi.org/10.1007/JHEP12(2021)188}{\emph{JHEP} {\bf 12} (2021)
  188}, [\href{http://arxiv.org/abs/2110.10519}{{\tt 2110.10519}}].

\bibitem{Klebanov:1999tb}
I.~R. Klebanov and E.~Witten, \emph{{AdS / CFT correspondence and symmetry
  breaking}},
  \href{http://dx.doi.org/10.1016/S0550-3213(99)00387-9}{\emph{Nucl. Phys. B}
  {\bf 556} (1999) 89--114}, [\href{http://arxiv.org/abs/hep-th/9905104}{{\tt
  hep-th/9905104}}].

\bibitem{Faulkner:2009wj}
T.~Faulkner, H.~Liu, J.~McGreevy and D.~Vegh, \emph{{Emergent quantum
  criticality, Fermi surfaces, and AdS(2)}},
  \href{http://dx.doi.org/10.1103/PhysRevD.83.125002}{\emph{Phys.Rev.} {\bf
  D83} (2011) 125002}, [\href{http://arxiv.org/abs/0907.2694}{{\tt
  0907.2694}}].

\bibitem{Birmingham:2001pj}
D.~Birmingham, I.~Sachs and S.~N. Solodukhin, \emph{{Conformal field theory
  interpretation of black hole quasinormal modes}},
  \href{http://dx.doi.org/10.1103/PhysRevLett.88.151301}{\emph{Phys. Rev.
  Lett.} {\bf 88} (2002) 151301},
  [\href{http://arxiv.org/abs/hep-th/0112055}{{\tt hep-th/0112055}}].

\bibitem{Natsuume:2019xcy}
M.~Natsuume and T.~Okamura, \emph{{Nonuniqueness of Green{\textquoteright}s
  functions at special points}},
  \href{http://dx.doi.org/10.1007/JHEP12(2019)139}{\emph{JHEP} {\bf 12} (2019)
  139}, [\href{http://arxiv.org/abs/1905.12015}{{\tt 1905.12015}}].

\bibitem{Camargo:2022rnt}
H.~A. Camargo, V.~Jahnke, K.-Y. Kim and M.~Nishida, \emph{{Krylov complexity in
  free and interacting scalar field theories with bounded power spectrum}},
  \href{http://dx.doi.org/10.1007/JHEP05(2023)226}{\emph{JHEP} {\bf 05} (2023)
  226}, [\href{http://arxiv.org/abs/2212.14702}{{\tt 2212.14702}}].

\bibitem{Banados:1992wn}
M.~Banados, C.~Teitelboim and J.~Zanelli, \emph{{The Black hole in
  three-dimensional space-time}},
  \href{http://dx.doi.org/10.1103/PhysRevLett.69.1849}{\emph{Phys. Rev. Lett.}
  {\bf 69} (1992) 1849--1851}, [\href{http://arxiv.org/abs/hep-th/9204099}{{\tt
  hep-th/9204099}}].

\bibitem{Koekoek_2010}
R.~Koekoek, P.~A. Lesky and R.~F. Swarttouw, \emph{Hypergeometric Orthogonal
  Polynomials and Their q-Analogues}.
\newblock Springer Berlin Heidelberg, 2010,
  \href{http://dx.doi.org/10.1007/978-3-642-05014-5}{10.1007/978-3-642-05014-5}.

\bibitem{Son:2002sd}
D.~T. Son and A.~O. Starinets, \emph{{Minkowski space correlators in AdS / CFT
  correspondence: Recipe and applications}}, {\emph{JHEP} {\bf 0209} (2002)
  042}, [\href{http://arxiv.org/abs/hep-th/0205051}{{\tt hep-th/0205051}}].

\bibitem{Natsuume:2020snz}
M.~Natsuume and T.~Okamura, \emph{{Pole-skipping and zero temperature}},
  \href{http://dx.doi.org/10.1103/PhysRevD.103.066017}{\emph{Phys. Rev. D} {\bf
  103} (2021) 066017}, [\href{http://arxiv.org/abs/2011.10093}{{\tt
  2011.10093}}].

\bibitem{Groenevelt:aa}
W.~Groenevelt, E.~Koelink and H.~Rosengren, \emph{Continuous hahn functions as
  clebsch-gordan coefficients}, {\emph{Theory and applications of special
  functions, 221--284, Dev. Math., 13, Springer, New York, 2005} },
  [\href{http://arxiv.org/abs/math/0302251}{{\tt math/0302251}}].

\bibitem{Muck:2022xfc}
W.~M{\"u}ck and Y.~Yang, \emph{{Krylov complexity and orthogonal polynomials}},
  \href{http://dx.doi.org/10.1016/j.nuclphysb.2022.115948}{\emph{Nucl. Phys. B}
  {\bf 984} (2022) 115948}, [\href{http://arxiv.org/abs/2205.12815}{{\tt
  2205.12815}}].

\end{thebibliography}
\bibliographystyle{JHEP}

\providecommand{\href}[2]{#2}\begingroup\raggedright\endgroup

\end{document}